%% file: paper_main_arxiv2.tex
\documentclass[journal]{IEEEtran}
\usepackage{cite}
\usepackage{url}
\usepackage{subcaption}
\usepackage[absolute]{textpos}

\usepackage{amsmath}

\newcommand{\copyrightstatement}{
    \begin{textblock}{15}(0.5,0.7)    
         \noindent
         \centering
         \textblockcolour{white}
         \footnotesize
         PREPRINT \copyright 2023 IEEE. Personal use of this material is permitted. Permission from IEEE must be obtained for all other uses, in any current or future media, including reprinting/republishing this material for advertising or promotional purposes, creating new collective works, for resale or redistribution to servers or lists, or reuse of any copyrighted component of this work in other works
    \end{textblock}}

\usepackage{psfrag}

\usepackage{graphicx}
\usepackage{xcolor}

\newcommand{\CCH}[1]{{\textcolor{black}{#1}}}
\newcommand{\CCHT}[1]{{\textcolor{black}{#1}}}
\newcommand{\BD}[0]{{\Delta^\mathrm{BD}}}

\newcommand{\BDsub}[0]{\Delta^\mathrm{BD}_{\mathrm{sub}}}
\newcommand{\BDall}[0]{\Delta^\mathrm{BD}_{\mathrm{all}}}

\begin{document}
%
\title{The Bj{\o}ntegaard Bible \\ Why your Way of Comparing Video Codecs May Be Wrong}
%
%
%


\author{Christian Herglotz, Hannah Och, Anna Meyer, Geetha Ramasubbu,  Lena Eicherm\"uller, \\ Matthias Kr\"anzler, Fabian Brand,  Kristian Fischer, Dat Thanh Nguyen, Andy Regensky, Andr\'e Kaup
\thanks{Christian Herglotz, Hannah Och, Anna Meyer, Geetha Ramasubbu,  Lena Eicherm\"uller, Matthias Kr\"anzler, Fabian Brand,  Kristian Fischer, Dat Thanh Nguyen, Andy Regensky, and Andr\'e Kaup are with the Chair of Multimedia Communications and Signal Processing, Friedrich-Alexander University Erlangen-N\"urnberg (FAU), Erlangen, Germany. e-mail: \{ firstname.lastname\}@fau.de}
\thanks{Manuscript accepted December 10th 2023.}}

%

\maketitle

\copyrightstatement

\begin{abstract}
In this paper, we provide an in-depth assessment on the Bj{\o}ntegaard Delta. We construct a large data set of video compression performance comparisons  using a diverse set of metrics including PSNR, VMAF, bitrate, and processing energies. These metrics are evaluated for visual data types such as classic perspective video, $\mathbf{360^\circ}$ video, point clouds, and screen content. As compression technology, we consider multiple hybrid video codecs as well as state-of-the-art neural network based compression methods. Using additional supporting points in-between standard points defined by parameters such as the quantization parameter, we assess the interpolation error of the Bj{\o}ntegaard-Delta (BD) calculus and its impact on the final BD value. From the analysis, we find that the BD calculus is most accurate in the standard application of rate-distortion comparisons with mean errors below $\mathbf{0.5}$ percentage points. For other applications \CCH{ and special cases, e.g., VMAF quality, energy considerations, or inter-codec comparisons}, the errors are higher (up to $\mathbf{5}$ percentage points), but can be \CCH{halved by using }a higher number of supporting points. We finally come up with recommendations on how to use the BD calculus such that the validity of the resulting BD-values is maximized. Main recommendations are \CCH{as follows:  First, relative curve differences should be plotted and analyzed. Second, the logarithmic domain should be used for saturating metrics such as SSIM and VMAF. Third, BD values below a certain threshold indicated by the subset error should not be used to draw recommendations. Fourth, using two supporting points is sufficient to obtain rough performance estimates. } 
\end{abstract}


%
\IEEEpeerreviewmaketitle

 

\section{Introduction}
%
%
%
%
\IEEEPARstart{V}{ideo} compression research has been a highly active and fruitful field of science in the past decades. Driven by more and more demand for consumer video applications, industry and academia has managed to develop a large variety of sophisticated and efficient compression techniques reducing transmission bitrates while keeping the visual quality of videos at a high level. 
Within the past two decades, various video codec specifications were standardized by organizations such as the Moving Pictures Experts Group (MPEG) with the codecs H.264/AVC \cite{ITU_H.264}, HEVC \cite{ITU_HEVC}, VVC \cite{ITU_VVC} or the Alliance for Open Media (AOM) and its predecessors with codecs such as VP8 \cite{VP8_spec}, VP9 \cite{VP9_spec}, and AV1 \cite{AV1_spec}. Currently, novel compression methods based on deep learning are investigated \cite{Balle18, Minnen18}, promising to further increase the compression efficiency of future codecs. For all these compression methods, the target was to reduce the bitrate as much as possible while keeping a high visual quality. 

Simultaneously, in particular throughout the last decade, new applications, use-cases, and target metrics for video compression appeared. First of all, it was noticed that classic visual quality metrics such as the peak signal-to-noise ratio (PSNR) have an imperfect correlation with subjective quality, i.e., the \CCHT{visual quality perceived by} human observers. As a consequence, a large variety of new metrics was developed, e.g., the structural similarity (SSIM) metric \cite{Wang04} or video multi-method assessment fusion (VMAF) \cite{VMAF}. For screen content, dedicated metrics were developed and analyzed \cite{Ni2017,Ni2018}. With the rise of machine learning techniques for tasks such as object detection or object segmentation, it was found that the detection accuracy in terms of the mean average precision~(mAP) is of higher importance than the visual quality \cite{Fischer20c, Fischer20b}.  

In another research direction, it was found that the energy consumption of video decoders and encoders plays an important role in portable devices and, due to billions of end users worldwide, also in terms of greenhouse gas emissions caused by electric power consumption \cite{ShiftFull19}. Hence, researchers tried to monitor and reduce the energy consumption of such devices \cite{Herglotz19,Khernache21,He13,Mercat17}. 

Finally, due to advancements in technology and new kinds of applications in, e.g., virtual or augmented reality (VR / AR), new kinds of visual data such as point clouds (PCs) \cite{Liu19,Li20} or $360^\circ$-videos \cite{Budagavi15,Duanmu18,Jamali19a} drew more and more attention in the research community. For these data types, new quality metrics had to be defined allowing to compare the performance of compression algorithms \cite{dmetric,Xiu17}. 

All of these compression technologies have one commonality: they compress visual data by allowing to introduce distortions (also called lossy compression) by jointly optimizing at least two target metrics. In classic video compression, these two metrics are the objectively measurable distortion and the bitrate. 
In theory, these two metrics cannot both be optimal at the same time, which is a consequence from the rate-distortion (RD) theory \cite{Berger71} that states that a lower distortion inevitably leads to a higher bitrate. 

Consequently, it is impossible to develop one codec \CCHT{that minimizes} both metrics, rate $R$ and distortion $D$. We always have to find a compromise between the two metrics. As a result, the concept of rate-distortion optimization (RDO) was developed \cite{Sullivan98}, which is a major cornerstone of all modern hybrid video codecs. Briefly speaking, RDO is a way to decide for coding tools and methods which lead to a good \CCHT{trade-off} between rate and distortion. Mathematically, this is achieved by minimizing a cost function 
\begin{equation}
\min J = D + \lambda	\cdot R, 
\end{equation}
where $\lambda$ is a Lagrange multiplier that can be freely chosen. In general, a large $\lambda$ leads to a low rate at high distortions, and a small $\lambda$ to high rates at low distortions. 

Consequently, the compression performance of codecs is not compared in terms of a single metric, but in terms of the relation between two metrics that can be connected by a curve. For rate-distortion considerations, this curve can be represented mathematically by the function $R = R(D)$. 

\begin{figure}
\psfrag{006}[tr][t]{ Rate in kbps}%
\psfrag{007}[bc][bc]{ PSNR in dB}%
\psfrag{000}[r][l]{ $1$}%
\psfrag{001}[r][l]{ $10$}%
\psfrag{002}[rc][rc]{ $34$}%
\psfrag{003}[rc][rc]{ $36$}%
\psfrag{004}[rc][rc]{ $38$}%
\psfrag{005}[rc][rc]{ $40$}%
\psfrag{Codec Aaaa}[l][l]{\small{Codec A}}%
\psfrag{Codec B}[l][l]{\small{Codec B}}%
\includegraphics[width=0.5\textwidth]{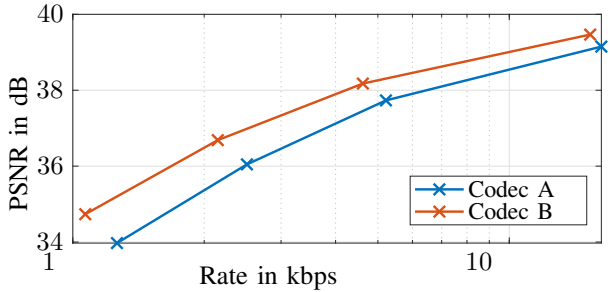}
\caption{Example for operational rate-distortion points (x-markers) of two codecs A (blue) and B (red) including a linear interpolation (solid lines). } 
\label{fig:RDexample}
\end{figure}

Hence, the most accurate way to compare the performance of two video codecs is comparing the rate-distortion behavior in a two-dimensional representation. An example for linearly interpolated curves with four supporting points of two codecs A (blue) and B (red) is illustrated in Fig.~\ref{fig:RDexample}. 
We can see that all supporting points of the red curve are located top-left of the blue points. As we desire a low bitrate and a high PSNR (i.e. a low distortion), we can conclude that the compression performance of codec B in terms of rate and distortion is superior to the compression performance of codec A. 

Due to restrictions in practical implementations, an encoder implementation cannot generate each position on a continuous curve, but only discrete operational points (x-markers in Fig.~\ref{fig:RDexample}). For better visibility in rate-distortion plots,  it is hence common to interpolate the discrete points by a continuous line. 

Researchers found that comparing two such curves visually is cumbersome. A metric simplifying this performance comparison and showing the performance difference in a single numerical value was desired. As a consequence, the  Bj{\o}ntegaard-Delta (BD) metric was developed \cite{Bjonte01} that calculates an average difference between two curves interpolating the distinct operating points of two codecs. 

After adoption of this metric for the use in RD comparisons, researchers decided to also use this calculus for other use-cases including different data types and performance metrics.  
Interestingly, the BD metric was adopted by researchers (including ourselves) without evaluating whether the BD calculus leads to representative and accurate results in use-cases different from classic RD comparisons. Also, a self-contained and understandable mathematical description of all algorithmic steps in the BD calculus was never provided. To this end, this paper aims at filling the gap and providing the following contributions: 
\begin{itemize}
\item a full mathematical description of calculation methods to derive the BD value,
\item an overview on existing implementations and their algorithmic properties, 
\item a list of current BD use cases,
\item an analysis on the accuracy of the interpolation algorithm using intermediate supporting points, 
\item an analysis on \CCH{the error imposed on the BD metric caused by subsampled} supporting points, 
\item recommendations for the correct and reliable usage of the BD metric in future research. 
\end{itemize}

This paper is organized as follows. First, Section~\ref{sec:lit} gives an overview on related literature targeting the BD calculus. Afterwards, Section~\ref{sec:bjonteCalc} shows the mathematical formulation and points out differences in existing implementations. Then, Section~\ref{sec:data sets} summarizes our data sets and experimental conditions. Section~\ref{sec:eval} evaluates the interpolation accuracy and the impact of the interpolation error on the final BD value. Finally, Section~\ref{sec:recs} recommends best-practices for using the BD and Section~\ref{sec:concl} concludes this paper.

\section{Literature Review}
\label{sec:lit}
The very first implementation of the BD calculus was introduced in a standardization document written by Gisle Bj{\o}ntegaard for a 2001 Video Coding Experts Group (VCEG) meeting \cite{Bjonte01}. Since then, there have been numerous implementations, studies, and standardization activities dealing with the BD's concept and its intricacies. An Excel add-in was proposed in \cite{VCEG-AE07}, implementing the calculus for practical application. Afterwards, it was proposed to separate the range of considered bitrate levels into high-rate and low-rate areas \cite{VCEG-AI11}. Furthermore, it was observed that it is important to consider the overlap of the two performance curves, which resulted in a reliability metric based on the overlap ratio \cite{VCEG_AL22}. 

Later, with the standardization of the High-Efficiency Video Coding (HEVC) codec, more studies and implementations were considered. In \cite{JCTVC-A031}, a tool evaluation test suit was proposed within an Excel implementation, which uses the BD calculus. This calculus was eventually used in the common test conditions (CTCs) for HEVC \cite{Bossen13}. 

Afterwards, the initial interpolation algorithm, polynomial fitting (polyfit)\footnote{Note that in the case of four supporting points, polyfit is equivalent to cubic spline interpolation (CSI).}, was replaced by piecewise-cubic hermite interpolation polynomial (PCHIP) \cite{JVET_Q0286,JVET_R2016}, which was found to yield more stable results.  PCHIP was also used for various other standardization efforts, including, e.g., VVC \cite{JVET_T2010}, $360^\circ$-degree video coding in VVC \cite{JVET_E1030}, AOM \cite{AOM_CTC21}, and video-based point cloud compression \cite{Schwarz18}. Next to the change of the interpolation algorithm, also the number of supporting points was increased as, e.g., \CCHT{in the latest CTCs by AOM} using six supporting points \cite{AOM_CTC21}. As a summarizing document, the ITU-T published working practices on how to use the BD calculus in video compression evaluations \cite{Strom21}. Here, typical problems were mentioned, which include unstable interpolation algorithms (in particular CSI/polyfit), little overlap of curves, and intersecting curves. 

Other works focused on extending or changing the BD calculus. For example, in \cite{JVET_H0030}, it was proposed to extrapolate the curves so that the overlapping interval is increased. For the application of subjective evaluation, it was proposed to use a logistic function to fit the rate-distortion curves in an approach called Subjective Comparison of ENcoders based on fitted Curves (SCENIC) \cite{Hanhart14}. \CCH{In a later work, it was found that SCENIC indeed outperforms PCHIP in case of subjective quality rating and the use of four supporting points \cite{Brand23}.} Furthermore, it was proposed to extend curve fitting to surface fitting in order to consider rate-distortion-complexity performance \cite{Li10} or rate-rate-distortion performance in the case of two-layer video compression \cite{Hanhart15}. 

Finally, some properties of the BD calculus were investigated in various papers and online publications. In \cite{Zimichev2022}, the difference between interpolation functions, namely polyfit and PCHIP, was discussed. In \cite{Barman22}, different implementations of the BD calculations were compared. Finally, in \cite{Herglotz22b}, the accuracy of the interpolated curves with respect to the true intermediate RD points was discussed in detail, which revealed that CSI/polyfit should not be used and that Akima interpolation is slightly more accurate than PCHIP.


A list of relevant and publicly available implementations of the BD calculus for different programming languages known to us is given in Table~\ref{tab:BDimpls}. The table also includes information on algorithmic details on the BD implementations such as the basis of the logarithm and the interpolation algorithm, which are all not explicitly defined in the original document \cite{Bjonte01}. For example, the transformation of the bitrate from the absolute to the logarithmic domain is sometimes performed with the natural logarithm ($\mathrm{ln}$), and sometimes with the logarithm to the basis $10$ ($\mathrm{log}10$). In addition, the number of allowed input parameters can differ. The columns ``BDR'' and ``BDPSNR'' indicate whether the implementation can calculate \CCHT{mean} relative bitrate differences and mean absolute distortion differences, respectively. The final column (CTC) indicates whether the implementation was used in common test conditions \CCHT{by standardization bodies}. 

\begin{table*}[t]
\caption{Publicly available implementations of the BD calculus, programming language, and important properties. VBA stands for Visual Basic for Applications (programming language used within Excel). Note that many of the easily accessible implementations use polyfit for interpolation, which can reportedly lead to unstable results \cite{Strom21, Herglotz22b}. }
\centering
\begin{tabular}{l|l|l|r|c|c|l|l|c}
\hline
Source & Name & Language          & No. params            & BDR & BDPSNR & Log & Interp. Algorithm & CTC \\
\hline
 \cite{googleCompCodec} & Google's Compare Codecs & Python & $\infty$ & x & x & $\ln$ & Polyfit & - \\
 \cite{VMAF_GITHUB} & Netflix VMAF & Python & $\infty$ & x & - & $\mathrm{log}10$ & PCHIP & - \\
 \cite{MatlabBD} & Bj. metric calculation& Matlab & $\infty$ & x & x & $\ln$ & Polyfit & -\\
 \cite{BDCPP} &Bj. Metric impl. for C++17 (or later)& C++ &  $\infty$ &  x & x & $\ln$ & Polyfit & -\\
 \cite{BDETRO}& ETRO's Bj. Metric impl. for Excel & Excel (VBA) &  $\infty$ &  x & x & $\ln$ & Polyfit & -\\
 \cite{BD_TS} & Bj. Metric impl. for JavaScript/TypeScipt & TypeScript &  $\infty$ &  x & x & $\mathrm{log}10$ & PCHIP & -\\
 \cite{JVET_T2010} & JVET CTC&  Excel (VBA) & $4$ & x & - & $\mathrm{log}10$ & PCHIP & x\\
 \cite{Schwarz18} &CTC for V-PCC & Excel (VBA) & $\infty$ & x & - & $\mathrm{log}10$ & PCHIP & x\\
 \cite{BD_LMS} &Bj.-Delta Interpolation & Python, Matlab & $\infty$ & x & - & $\mathrm{log}10$ & CSI, PCHIP, Akima & -\\
 \hline
 \end{tabular}
\label{tab:BDimpls}
\vspace{-.3cm}
\end{table*}

We can see that there is an unexpected variety of implementations available. In this paper, we will discuss mathematical differences and their impact on BD values in detail and propose guidelines to avoid caveats when using different implementations. 

\section{The Bj{\o}ntegaard Calculation}
\label{sec:bjonteCalc}

\begin{figure}
\psfrag{014}[c][c]{$P_\mathrm{dep}$}%
\psfrag{015}[bc][bc]{$P_\mathrm{ind}$}%
\psfrag{000}[tc][tc]{ }%
\psfrag{001}[r][ct]{ $0$}%
\psfrag{002}[r][ct]{ $5$}%
\psfrag{003}[r][ct]{ $10$}%
\psfrag{004}[r][ct]{ $15$}%
\psfrag{005}[r][ct]{ $20$}%
\psfrag{006}[rc][rc]{ $33$}%
\psfrag{007}[rc][rc]{ $34$}%
\psfrag{008}[rc][rc]{ $35$}%
\psfrag{009}[rc][rc]{ $36$}%
\psfrag{010}[rc][rc]{ $37$}%
\psfrag{011}[rc][rc]{ $38$}%
\psfrag{012}[rc][rc]{ $39$}%
\psfrag{013}[rc][rc]{ $40$}%
\psfrag{Codec A}[l][l]{\small{Support points A}}%
\psfrag{Interp. A}[l][l]{\small{Interpolation A}}%
\psfrag{Codec B}[l][l]{\small{Support points B}}%
\psfrag{Interp. B}[l][l]{\small{Interpolation B}}%
\psfrag{bounds}[l][l]{\small{Integration bounds}}%
\psfrag{Integration Area}[l][l]{\small{Integration Area}}%
\includegraphics[width=0.5\textwidth]{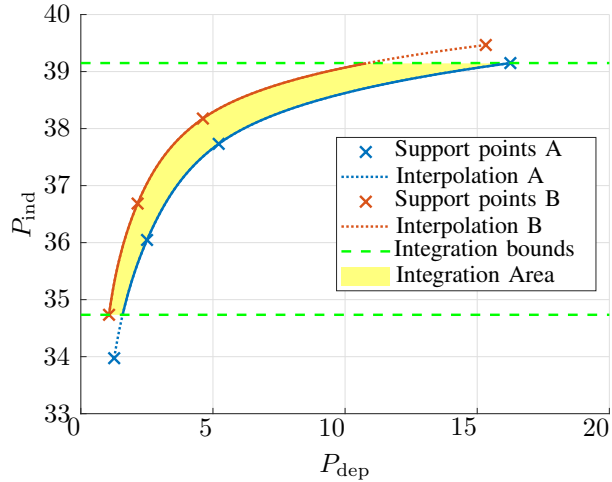}
\caption{Visualization of the BD calculus. A set of input points (support points, x-markers) is determined for two codecs. In many applications, the independent variable $P_\mathrm{ind}$ corresponds to a quality metric such as PSNR, and the dependent variable $P_\mathrm{dep}$ to a bitrate in kbps. For each codec, an interpolation is calculated (solid and dotted lines). Then, the overlapping interval of the independent variable is determined (dashed green lines). Finally, the difference between the two curves is integrated over the overlapping interval (yellow area). } 
\label{fig:BDvisual}
\end{figure}

The BD-calculations were first introduced in \cite{Bjonte01} and are used to compare the performance difference between two encoder implementations or encoder configurations $k\in\{\mathrm{A},\mathrm{B}\}$. The performance is usually measured and averaged for a predefined set of video sequences. For simplicity, we discuss the calculations for a single sequence. A visualization of the main steps is illustrated in Fig.~\ref{fig:BDvisual}. 

First, the two encoders or encoder configurations $\mathrm{A}$  and $\mathrm{B}$ are defined. For example, $\mathrm{A}$  and $\mathrm{B}$ could represent two different codecs, two different encoders for the same codec, or the same encoder where a certain single coding tool is switched on and off. 

Afterwards, a set of supporting points is chosen for both $\mathrm{A}$ and $\mathrm{B}$. In classic standardization, a supporting point is defined by the quantization parameter (QP), which is used to determine the trade-off between rate and distortion \cite{Sullivan98,JVET_T2010}. Often, $I=4$ supporting points are selected. Other works use similar parameters such as, e.g., the constant rate factor (crf) in certain encoder implementations \cite{Herglotz20b}. 

Then, two performance metrics (PMs) are selected, which are classically the distortion $D$ in terms of the peak signal-to-noise ratio (PSNR) and the bitrate $R$. 
Next, the input video is encoded with both encoder configurations $k\in\{\mathrm{A},\mathrm{B}\}$ and the selected set of supporting points indexed by $i$. The two PMs for the resulting eight compressed videos are then used for subsequent calculations.  
The PM values must be arranged in monotonic order. 


Thereafter, a functional relation is determined between the two PMs. For this, it is decided which of the PMs is used as the dependent variable and which is used as the independent variable in the interpolated function. For example, in case of RD comparisons,  a function $R(D)$ or $D(R)$ is generated. In \cite{Bjonte01}, both variants are proposed. In practice, the most common approach is to define the rate as the dependent variable and the quality metric as the independent variable $R(D)$  \cite{JVET_T2010,Bossen13}, which we adopt in this paper. For the sake of generality and to show that other PMs could be used, we use the terms $P_\mathrm{ind}$ (e.g., the PSNR) and $P_\mathrm{dep}$ (e.g., the bitrate) for the independent and the dependent variable, respectively, in the following. 

In \cite{Bjonte01}, it was observed that the values of the bitrate (i.e., the dependent variable) could range multiple orders of magnitude. As a consequence, the bitrate was converted to the logarithmic domain by
\begin{equation}
p_\mathrm{dep}(k,i) = \log_{10}\big(P_\mathrm{dep}(k,i)\big), 
\label{eq:logR}
\end{equation}
where the lowercase $p$ represents the logarithmic domain. This step was performed to ensure that mean BD rate values are not biased towards higher bitrates \cite{Bjonte01}. Using the logarithm for the dependent variable was also adopted for other PMs replacing the bitrate, e.g., the decoding energy in \cite{Herglotz19,Kraenzler21}. Note that some implementations use the natural logarithm instead of the logarithm to the basis $10$ (see Table~\ref{tab:BDimpls}). However, theoretically, using a different basis in the calculation of the logarithm leads to the same BD value, because a different basis corresponds to a simple linear factor in the interpolation (see also Appendix C). 
Observed differences are hence caused by numerical inaccuracies due to floating point operations. 

Afterwards, piecewise polynomial curves are interpolated for both encoder configurations $k$, which use the positions of the independent variable $P_\mathrm{ind}(k,i)$ and the dependent variable $p_\mathrm{dep}(k,i)$ as supporting points. All $I-1$ pieces of the interpolated curve are represented by third order polynomials. The resulting piecewise polynomial curve can then be written as
\begin{equation}
\hat p_\mathrm{dep}(k,\phi) 
= \begin{cases} \hat p_{\mathrm{dep},1}(k,\phi)= a_{k,1} + b_{k,1}\cdot \phi\\
 \quad  + c_{k,1}\cdot \phi
 ^2 + d_{k,1}\cdot \phi
 ^3, \\
 \quad\quad  P_\mathrm{ind}(k,1)\le\phi<P_\mathrm{ind}(k,2)\\ 
 \quad\quad\vdots\\
 \hat p_{\mathrm{dep},i}(k,\phi)= a_{k,i} + b_{k,i}\cdot \phi\\ \quad 
 + c_{k,i}\cdot \phi
 ^2 + d_{k,i}\cdot \phi
 ^3, \\
 \quad\quad  P_\mathrm{ind}(k,i)\le\phi<P_\mathrm{ind}(k,i+1)\\
 \quad\quad\vdots\\
 \hat p_{\mathrm{dep},I-1}(k,\phi) = a_{k,I-1} + b_{k,I-1}\cdot \phi\\ \quad  
 + c_{k,I-1}\cdot \phi
 ^2 + d_{k,I-1}\cdot \phi
 ^3, \\
 \quad\quad  P_\mathrm{ind}(k,I-1)\le\phi<P_\mathrm{ind}(k,I),\\
  \end{cases}
\label{eq:BDpoly}
\end{equation}
where $\phi$ is an auxiliary variable representing the independent variable and the parameters $a_{k,i}$, $b_{k,i}$, $c_{k,i}$, and $d_{k,i}$ are derived by interpolation for both encoder configurations. The interpolation methods used in the literature, which are mainly cubic spline interpolation (CSI), piecewise cubic hermite polynomial interpolation (PCHIP), and Akima interpolation, are further explained and discussed in Appendix A.

Then, the difference between the two resulting interpolations is calculated by integration. The upper and lower bounds are selected as the overlapping part of the independent variable (e.g., the PSNR). The bounds of both sets of supporting points are calculated by  
\begin{align}
P_\mathrm{ind,low} = & \max\Big( \min\big(P_\mathrm{ind}(\mathrm{A},i)\big),\min\big(P_\mathrm{ind}(\mathrm{B},i)\big)\Big) \label{eq:lowBound} \\
P_\mathrm{ind,high} = & \min\Big( \max\big(P_\mathrm{ind}(\mathrm{A},i)\big),\max\big(P_\mathrm{ind}(\mathrm{B},i)\big)\Big)  \label{eq:highBound}. 
\end{align}

Finally, the average relative difference between the two interpolated curves, which we call the Bj{\o}ntegaard-Delta (BD) or $\BD$ in the following, is obtained by
\begin{equation}
\BD = 10^{\frac{1}{\Delta P_\mathrm{ind} } \int_{P_\mathrm{ind,low}}^{P_\mathrm{ind,high}}
 \hat p_\mathrm{dep}(\mathrm{A},P_\mathrm{ind}) 
 - \hat p_\mathrm{dep}(\mathrm{B},P_\mathrm{ind}) \mathrm{d}P_\mathrm{ind}} -1, 
 \label{eq:bdIntegration}
\end{equation}
with the integration range $\Delta P_\mathrm{ind} =  P_\mathrm{ind,high}-P_\mathrm{ind,low}$. \CCHT{$\BD$} describes the relative bitrate difference in the bounds of \eqref{eq:lowBound} and \eqref{eq:highBound} when encoded with  encoder $k=\mathrm{A}$ with respect to the reference encoder $k=\mathrm{B}$. Note that the mean difference (i.e. the integration) is calculated in the logarithmic domain, which puts an emphasis on small differences and reduces the impact of large differences when compared to calculating the mean difference in the non-logarithmic domain (see Appendix B for more details). Hence, the BD value can be interpreted as the mean relative difference averaged in the logarithmic domain. 

As a desirable side effect of integrating in the logarithmic domain, this formulation allows calculating the integration in a closed-form solution. The reason is that the integration is performed over a polynomial. In the other case, where the integration would be performed in the non-logarithmic domain ($\int 10^{\hat p_{\mathrm{dep}}(P_\mathrm{ind})} \mathrm{d}P_\mathrm{ind}$), no closed-form solution can be calculated due to the polynomial in the exponent. 

Mathematically, the definition of the BD calculus is clear and unique in most of its components. The only uncertainty is the choice of the interpolation method to determine $a_{k,i}$, $b_{k,i}$, $c_{k,i}$, and $d_{k,i}$ in Eq.~\eqref{eq:BDpoly}. In the past, several methods were proposed and used. In this work, we discuss three interpolation methods. Two methods were applied in practice (CSI, PCHIP) \cite{Bossen13,JVET_T2010} and one was claimed to outperform the former two in the literature (Akima) \cite{Herglotz22b}. 
 
It should be mentioned that all interpolation methods are constructed from third-order piecewise-cubic polynomials. As a consequence, for each piece of the piecewise polynomial, four coefficients need to be determined. Thus, each interpolated curve is described by $M=4\cdot(I-1)$ coefficients, such that in the case of $I=4$ supporting points, we need $M=12$ conditions.

\begin{figure}
\psfrag{013}[r][tc]{ $P_\mathrm{ind}$}%
\psfrag{014}[bc][bc]{ $P_\mathrm{dep}$}%
\psfrag{000}[r][ct]{ $1$}%
\psfrag{001}[r][ct]{ $2$}%
\psfrag{002}[r][ct]{ $3$}%
\psfrag{003}[r][ct]{ $4$}%
\psfrag{004}[r][ct]{ $5$}%
\psfrag{005}[r][ct]{ $6$}%
\psfrag{006}[r][l]{ $7$}%
\psfrag{007}[rc][rc]{ $0$}%
\psfrag{008}[rc][rc]{ $2$}%
\psfrag{009}[rc][rc]{ $4$}%
\psfrag{010}[rc][rc]{ $6$}%
\psfrag{011}[rc][rc]{ $8$}%
\psfrag{012}[rc][rc]{ $10$}%
\psfrag{Supp poooooooo}[l][l]{{Support points}}%
\psfrag{Spline}[l][l]{ CSI}%
\psfrag{PCHIP}[l][l]{PCHIP}%
\psfrag{Akima}[l][l]{ Akima}%
\includegraphics[width=0.5\textwidth]{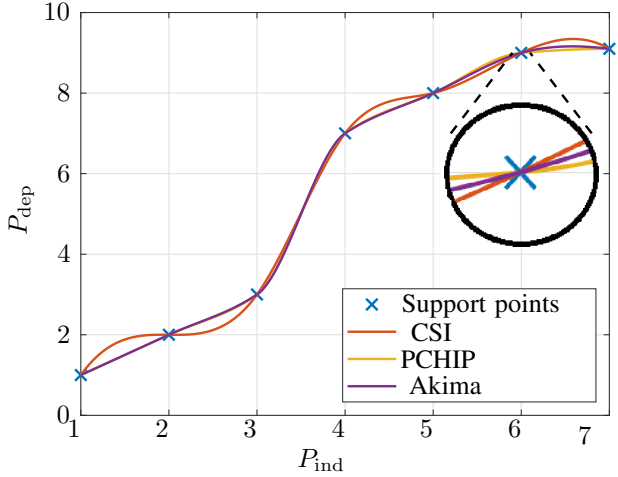}
\caption{Example for support points (blue x-markers) and interpolated curves using CSI (red), PCHIP (yellow), and Akima (purple). The support points are located at $P_\mathrm{ind}=\{1, 2, ..., 7\}$ and at $P_\mathrm{dep}=\{1, 2, 3, 7, 8, 9, 9.1\}$. }
\label{fig:exampleInterpCurves}
\end{figure}

Exact mathematical definitions and explanations for the three considered interpolation methods CSI, PCHIP, and Akima are provided in \CCHT{Appendix A}. From those explanations, we can find the following observations, which we illustrate in Fig.~\ref{fig:exampleInterpCurves}.  
\begin{itemize}
\item CSI can lead to overshoots (Runge's phenomenon \cite{Runge1901}). In Fig.~\ref{fig:exampleInterpCurves}, the interpolated curve shows overshoots when the slopes change, e.g., between $P_\mathrm{ind}=1$ and $P_\mathrm{ind}=3$. Hence, CSI should not be used in practice.
\item In PCHIP interpolation, derivatives at supporting points are biased towards small values. In Fig.~\ref{fig:exampleInterpCurves}, this can be observed between $P_\mathrm{ind}\in[6,7]$, where the derivative is the smallest of all interpolation methods and consequently, the curve is the lowest. 
\item In Akima interpolation, derivatives at supporting points are biased towards homogeneous values in the local region (adjacent supporting points). In Fig.~\ref{fig:exampleInterpCurves}, this can be observed at $P_\mathrm{ind}=6$, where the derivative is close to the derivative at $P_\mathrm{ind}=5$. 
\end{itemize}

\section{data sets}
 \label{sec:data sets}
 
With the data sets studied in this paper, our goal is to cover a wide variety of applications and special cases used in practice. This chapter provides an overview on the used data sets with a summary in Table~\ref{tab:datasets}. As will be shown in Section~\ref{sec:eval}, our evaluation mainly relies on the use of intermediate supporting points. That means that next to performance points generated by standard conditions (e.g., QPs $22$, $27$, $32$, $37$ in JVET standardizations \cite{JVET_T2010}), we additionally investigate the performance at intermediate supporting points. 

\begin{table*}[t]
\renewcommand{\arraystretch}{1.3}
\caption{Overview on tested data sets. The ID helps to identify the data set, the type indicates the data type, and the size the number of input images or sequences. \CCHT{pQS is the \texttt{positionQuantisationScale} from G-PCC v14.} `P' and `b' are short for PSNR and bitrate, respectively. The reference encoder configuration is highlighted in bold font. }
\label{tab:datasets}
\vspace{-0.3cm}
\begin{center}
\resizebox{\textwidth}{!}{ 
\begin{tabular}{c|l|c|r|c|l|l}
\hline
ID & Type & Source & Size & Basic supporting points &  Performance metrics (PMs) & Encoder configurations\\
\hline
RAT & Video & \cite{JVET_T2010} & $23$ &$\mathrm{QP}\in\{22,27,32,37\}$ & P, b, VMAF, decoding energy & \textbf{VVenC} RA; ISP off; SAO off; LMCS off; \\ 
 & & & & & & \quad BDOF off; ALF off  \\ 
LDT & Video & \cite{JVET_T2010} & $23$ &$\mathrm{QP}\in\{22,27,32,37\}$ & P, b, VMAF & \textbf{VVenC} LDB; ISP off; SAO off; LMCS off; \\ 
 & & & & & & \quad  BDOF off; ALF off \\
CC & Video & \cite{JVET_T2010} & $23$ &$\mathrm{QP}\in\{22,27,32,37\}$ & P, b, VMAF & \textbf{VTM-17.2}; HM-16.26 \\ 
ENC & Video & \cite {JVET_T2010} &22 &$\mathrm{CRF}\in\{18,23,28,33\}$& P, b, VMAF, SSIM, enc. energy \& time &  \textbf{x265} and x264 with presets \textbf{veryfast},    \\
& & & & & & \quad medium, {and slower}\\
AOM & Video & \cite{AOM_CTC21} & $8$ & $\mathrm{QP}\in\{17,22,27,32,37,42\}$ & P, b, SSIM, MS-SSIM, VMAF, enc. time & \textbf{VP9} and AV1 with presets \textbf{2}, 6, and 9  \\
NNV & Video & \cite{Li2022} & $37$ &$\mathrm{QS}\in\{1.541, 1.083, 0.729, 0.5\}$ & P, b, MS-SSIM, VMAF & \textbf{DMC-32}; VTM-17.2; HM-16.25 \\ 
SCC & Image & \cite{Yang2021} & $200$ &$\mathrm{QP}\in\{22,27,32,37\}$ & P, b, SSIM, ESIM, GFM, GMSD & \textbf{VTM-17.2-SC} IBC, PLT, and IBC+PLT off;  \\
& & & & & & \quad HM16.21-SCM8.8-SC\\
VCM & Image & \cite{Cordts16} & 500 & $\mathrm{QP}\in\{12,17,22,27,32,37\}$ & P, b, VMAF, SSIM, mAP, wAP, mIoU & \textbf{VTM-10.0}; HM-16.18  \\
360 & Video & \cite{JVET_E1030} & $10$ & $\mathrm{QP}\in\{22,27,32,37\}$ & P, b, (WS, S, CPP, Viewport)-PSNR & \textbf{ERP format}; HEC format  \\
PCA & PC-Video & \cite{8i} & $10$ &  $\mathrm{QP}\in\{22,28,34,40,46,51\}$&PSNR (Y, Cb, Cr), bitrate& \textbf{G-PCC v14}; RAHT  \\
PCG & PC-Video & \cite{8i} & $10$ &  \CCHT{$\mathrm{pQS}\in\{0.03125, 0.0625,  $}&D1-PSNR, D2-PSNR, bitrate& \textbf{G-PCC v14 predictive}; non-predictive  \\
& & & & \CCHT{$ 0.125, 0.25, 0.5, 0.75\}$} & &  \\
\end{tabular}}
\end{center}
\vspace{-.6cm}
\end{table*}


For the main part of our evaluation, we focus on codec standardization. 
In standardizations, 
\input{datasets/Matthias/RAT.tex}
In the following, this dataset is separated into different encoder configurations and abbreviated by randomaccess (RA) tool test (RAT) and lowdelay\_B (LD) tool test (LDT), where energy measurements were only performed for RAT. 

Furthermore, we performed tests on various other datasets listed in Table~\ref{tab:datasets}. \CCH{These datasets comprise various visual data formats and corresponding  compression methods, which are commonly evaluated using the BD method. First, we consider codec comparisons (CC), where we compare the compression performance of two hybrid video codecs (VVC and HEVC). Second, we compare encoders at different presets (ENC). Third, we report results from standardizations in an alternative group, the Alliance for Open Media (AOM) with the two codecs AV1 and VP9. Fourth, we assess the compression performance of neural-network-based video compression (NNV) solutions. }

\CCH{We assess compression methods for visual data targeting different applications. First, we consider screen content coding (SCC), where we investigate dedicated metrics that have a stronger correlation to the perceived visual quality than traditional metrics such as PSNR or SSIM. Second, we consider the object detection accuracy in video-coding-for-machines (VCM) scenarios. Third, we assess the   compression performance of $360^\circ$ video compression methods using dedicated metrics such as the weighted sphere PSNR (WS-PSNR). Finally, we consider performance metrics used in point cloud compression using both the attributes (PCA) and the geometry (PCG). More detailed explanations on all datasets are provided in the Appendix D. }

\CCH{In the following, we mainly focus our investigations on the RAT dataset.  Results from other datasets are reported when they differ significantly or when they show striking properties. }

\section{Evaluation}
\label{sec:eval}

In this section, we present an evaluation of the BD calculus and assess its accuracy. First, we concentrate on errors of the interpolated curves, similar to the results presented in \cite{Herglotz22b} in Subsection~\ref{sec:interAcc}. Second, we propose a novel assessment technique called relative curve difference (RCD) in Subsection~\ref{sec:RCD}. Motivated by this visualization, we evaluate the impact of the number of supporting points on the accuracy of the final $\BD$ value in Subsection~\ref{sec:BDAcc}. 
Due to conciseness reasons, in this paper, we focus on values on the RAT dataset. 
The error metrics for all data sets and performance metrics listed in Table~\ref{tab:datasets} can be found in Appendix E.

\subsection{Interpolation Accuracy}
\label{sec:interAcc}


To assess the interpolation accuracy, we adopt the concept introduced in \cite{Herglotz22b}, where the interpolation error is evaluated on all available performance points between the standard supporting points. For example, in the RAT data set (JVET standardizations \cite{JVET_T2010}), we calculate the interpolated curve using the four supporting points $\mathrm{QP}\in\{22,27,32,37\}$ and evaluate the interpolated curve's accuracy on all points that can be reached in this QP range ($\mathrm{QP}\in \{22, 23, 24, ..., 37\}$). In the following, we will call the reduced set of QPs the ``subset points'' and the full set of points ``all points''. The mean relative interpolation error \CCH{(RIE)} for a single interpolated curve is thus calculated by  \cite{Herglotz22b}
\begin{equation}
\bar e^\mathrm{RIE} = \frac{1}{\left|\mathcal{P}\right|} \sum_{i\in \mathcal{P}} \frac{\left| 10^{\hat p_\mathrm{dep}(P_{\mathrm{ind},i})} - P_{\mathrm{dep},i}\right|}{P_{\mathrm{dep},i}},  \label{eq:RIE}
\end{equation}
where $\mathcal{P}$ is the set of all points, including the subset points. Furthermore, we assess the maximum interpolation error given by 
\begin{equation}
E_\mathrm{max}^\mathrm{RIE} = \max_{i\in \mathcal{P}}  \frac{\left| 10^{\hat p_\mathrm{dep}(P_{\mathrm{ind},i})} - P_{\mathrm{dep},i}\right|}{P_{\mathrm{dep},i}}.  \label{eq:maxE}
\end{equation}
Mean interpolation errors for all data sets and all tested performance metrics are listed in \CCHT{Appendix E}. 

\begin{figure*}
\psfrag{026}[bc][bc]{ Interpolation errors $e^\mathrm{RIE}$}%
\psfrag{000}[cb][ct]{ \small{\textit{PCHIP}}}%
\psfrag{001}[rt][cb]{ \rotatebox{30}{{PSNR}-bitrate}}%
\psfrag{002}[rb][ct]{ \small{\textit{Ak.}}}%
\psfrag{003}[cb][ct]{ \small{\textit{PCHIP}}}%
\psfrag{004}[rt][cb]{ \rotatebox{30}{{PSNR}-time}}%
\psfrag{005}[rb][ct]{ \small{\textit{Ak.}}}%
\psfrag{006}[cb][ct]{ \small{\textit{PCHIP}}}%
\psfrag{007}[rt][cb]{ \rotatebox{30}{{PSNR}-dec. energy}}%
\psfrag{008}[rb][ct]{ \small{\textit{Ak.}}}%
\psfrag{009}[cb][ct]{ \small{\textit{PCHIP}}}%
\psfrag{010}[rt][cb]{ \rotatebox{30}{{VMAF}-bitrate}}%
\psfrag{011}[rb][ct]{ \small{\textit{Ak.}}}%
\psfrag{012}[cb][ct]{ \small{\textit{PCHIP}}}%
\psfrag{013}[rt][cb]{ \rotatebox{30}{{VMAF}-dec. energy}}%
\psfrag{014}[rb][ct]{ \small{\textit{Ak.}}}%
\psfrag{015}[cb][ct]{ \small{\textit{PCHIP}}}%
\psfrag{016}[rt][cb]{ \rotatebox{30}{{log VMAF}-bitrate}}%
\psfrag{017}[rb][ct]{ \small{\textit{Ak.}}}%
\psfrag{018}[cb][ct]{ \small{\textit{PCHIP}}}%
\psfrag{019}[rt][cb]{ \rotatebox{30}{log VMAF-dec. energy}}%
\psfrag{020}[rb][ct]{ \small{\textit{Ak.}}}%
\psfrag{021}[rc][rc]{ $0\%$}%
\psfrag{022}[rc][rc]{ $1\%$}%
\psfrag{023}[rc][rc]{ $2\%$}%
\psfrag{024}[rc][rc]{ $3\%$}%
\psfrag{025}[rc][rc]{ $4\%$}%
\includegraphics[width=\textwidth]{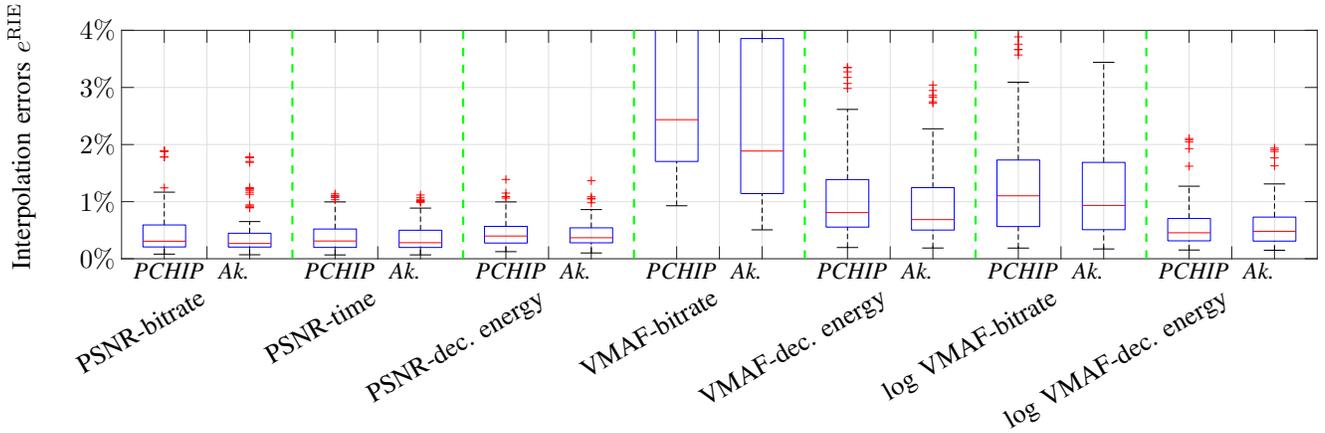}
\vspace{1.5cm}
\caption{\CCH{Box-whisker plots for the distribution of the interpolation error for the RAT dataset and all PM pairs. } }
\label{fig:boxplotInterpError}
\vspace{0cm}
\end{figure*}

\CCH{On top of the results reported in \cite{Herglotz22b}, which stated that Akima interpolation is more accurate than PCHIP, we perform an analysis on the distribution of interpolation errors. To this end, we collect all errors for PCHIP and Akima interpolation from the RAT dataset and generate a box-whisker plot on RIE statistics in Fig.~\ref{fig:boxplotInterpError}. }

\CCH{On the horizontal axis, each pair of boxes shows statistics for PCHIP (left) and Akima (right), respectively, for a pair of PMs (indicated below). The vertical axis represents the absolute value of the RIEs. The box'es boundaries indicate the $25$th and the $75$th percentile, the red line the median, the whiskers the boundaries of all values except for outliers, and the red crosses the outliers. }

\CCH{We can see that for the three leftmost pairs (PSNR-bitrate, PSNR-time, and PSNR-decoding energy), the overall RIEs are lowest. Only in exceptional cases, the RIEs are higher than $1\%$. For all these three PM pairs, more than $50\%$ of the RIEs are located below $0.5\%$. 
The four rightmost PM pairs show higher errors, indicating that using VMAF as a quality metric leads to less accurate interpolations. This behavior will be discussed in detail below. }

\CCH{Furthermore, we compare the RIE of PCHIP with the RIE of Akima interpolation. In general, for most pairs, Akima interpolation is slightly more accurate then PCHIP (e.g., boxes are located at lower values for PSNR-bitrate, VMAF-bitrate, VMAF-decoding energy), which we also observe for other datasets. This confirms the results from \cite{Herglotz22b}, where it was shown that Akima interpolation leads to more accurate interpolated curves. In Subsection~\ref{sec:BDAcc}, we will discuss the impact of this difference on actual $\BD$ values. Before, we will discuss two observations that should be considered when using the BD metric. }

\subsubsection{Saturating Quality Metrics}
\label{sec:satMetrics}
\begin{figure}
\psfrag{014}[rc][tc]{ Bitrate in Mbps}%
\psfrag{015}[bc][bc]{ VMAF Score}%
\psfrag{000}[tc][tc]{ }%
\psfrag{001}[r][ct]{ {$0$}}%
\psfrag{002}[r][ct]{ $2$}%
\psfrag{003}[r][ct]{ $4$}%
\psfrag{004}[r][ct]{ $6$}%
\psfrag{005}[r][ct]{ $8$}%
\psfrag{006}[r][ct]{ $10$}%
\psfrag{007}[rc][rc]{ $75$}%
\psfrag{008}[rc][rc]{ $80$}%
\psfrag{009}[rc][rc]{ $85$}%
\psfrag{010}[rc][rc]{ $90$}%
\psfrag{011}[rc][rc]{ $95$}%
\psfrag{012}[rc][rc]{ $100$}%
\psfrag{013}[rc][rc]{ }%
\psfrag{data5aaaaaa}[l][l]{\small All points}%
\psfrag{data2}[l][l]{\small Subset points}%
\psfrag{data3}[l][l]{\small VMAF}%
\psfrag{data4}[l][l]{\small log-VMAF}%
\includegraphics[width=0.5\textwidth]{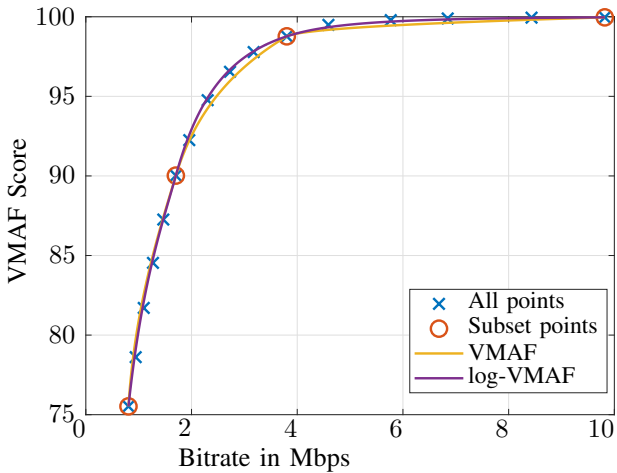}
\caption{ Subset points (o's), additional points (x'es) and interpolated curves (Akima interpolation) for the VMAF-bitrate performance of the RaceHorses sequence (class C), CC data set, VTM encoding.  For the yellow curve and the purple curve, interpolation is performed in the VMAF domain and in the log-VMAF domain as specified in Eq.~\eqref{eq:logVMAF}, respectively. } 
\label{fig:interpAccEx2}
\end{figure}

\CCH{Because of the observation that RIEs are high for VMAF, we discuss corresponding interpolated curves in detail. }Fig.~\ref{fig:interpAccEx2} shows an example for a VMAF-bitrate curve from the CC (codec comparison) data set.  
The curve shows that below a bitrate of $2\,$Mbps, the interpolation seems accurate (yellow curve hidden by purple curve). However, above $2\,$Mbps, the interpolation fails due to the saturation of the VMAF score, which results in considerable horizontal interpolation errors. This behavior could be observed for several metrics that show a saturation such as 
 SSIM, \CCH{mean average precision (mAP), and mean intersection over union (mIoU), where the latter two metrics are used in VCM tasks}. We conclude that when using non-logarithmic quality metrics, the BD calculus can lead to inaccurate interpolations.

Saturation was also observed in other studies using the SSIM metric \cite{Rassool17}. As a consequence, it was proposed to use a logarithmized SSIM metric in the decibel range as 
\begin{equation}
-10 \cdot \log_\mathrm{10}\left(1-\mathrm{SSIM}\right). 
\label{eq:logSSIM}
\end{equation}
As the VMAF metric often shows a similar behavior, we propose to apply \CCH{a similar concept to VMAF as} 
\begin{equation}
-10 \cdot \log_\mathrm{10}\left(1-\frac{\mathrm{VMAF}}{100}\right), 
\label{eq:logVMAF}
\end{equation}
\CCH{where we divide the VMAF score by $100$ to obtain a maximum value of one. }
\CCH{Statistics on RIEs for the case of VMAF are visualized in Fig.~\ref{fig:boxplotInterpError}.}   
We can see that from the interpolation point of view, RIEs for VMAF are significantly lower in the logarithmic domain \CCH{(around $2\%$ vs. around $1\%$)}. Also, the purple curve in Fig.~\ref{fig:interpAccEx2}, which is interpolated in the log-VMAF domain, shows that especially in the saturation region, the interpolation is more accurate. Furthermore, the logarithmic representation puts a higher weight on the part of the curves located at high VMAF values, which is desirable for high-quality applications. 

Concerning SSIM, we find that the \CCH{mean RIE} drops from $1.152\%$ to $0.617\%$ using Eq.~\eqref{eq:logSSIM} (SSIM-bitrate in ENC dataset, Appendix E). 
As such, we recommend to use the logarithmic representation for SSIM and VMAF in $\BD$ calculations. \CCH{In Section~\ref{sec:BDAcc}, we will discuss the impact of the logarithmic representation on the final $\BD$ values. }

For detection accuracy metrics in VCM scenarios (mIoU, mAP, or wAP), which also show saturation towards the detection precision of pristine data, we could not observe decreased interpolation errors \CCH{(e.g., $\bar e^\mathrm{RIE}$ of $2.746$ in the initial domain and $2.750\%$ in the logarithmic domain for VCM, high QP, Yolov5, wAP-bitrate). Hence,} we recommend to stick to the use of such metrics in the initial domain. \CCH{Further examples} can be found in Appendix E. The reason is that the metric's value of convergence is usually not the maximum value of the allowed range as in SSIM and VMAF ($1$ and $100$, respectively), but the detection precision of pristine data. A solution to this problem could be investigated in future research.

\subsubsection{Monotony of Supporting Points} 
\begin{figure}
\psfrag{data1aaaaaaaaa}[l][l]{\small All points}%
\psfrag{data2}[l][l]{\small Subset points}%
\psfrag{data3}[l][l]{\small Interpolation}%
\psfrag{019}[tc][tc]{  Bitrate in bits per pixel}%
\psfrag{020}[bc][bc]{ wAP}%
\psfrag{021}[tc][tc]{  Bitrate in bits per pixel}%
\psfrag{022}[bc][bc]{ mAP}%
\psfrag{000}[r][ct]{ $0$}%
\psfrag{001}[r][ct]{ $0.5$}%
\psfrag{002}[r][ct]{ $1$}%
\psfrag{009}[r][ct]{ $0$}%
\psfrag{010}[r][ct]{ $0.5$}%
\psfrag{011}[r][ct]{ $1$}%
\psfrag{003}[rc][rc]{ $0.34$}%
\psfrag{004}[rc][rc]{ }%
\psfrag{005}[rc][rc]{ $0.35$}%
\psfrag{006}[rc][rc]{ }%
\psfrag{007}[rc][rc]{ $0.36$}%
\psfrag{008}[rc][rc]{ }%
\psfrag{012}[rc][rc]{ }%
\psfrag{013}[rc][rc]{ $0.28$}%
\psfrag{014}[rc][rc]{ }%
\psfrag{015}[rc][rc]{ $0.29$}%
\psfrag{016}[rc][rc]{ }%
\psfrag{017}[rc][rc]{ $0.3$}%
\psfrag{018}[rc][rc]{ }%
\includegraphics[width=0.5\textwidth]{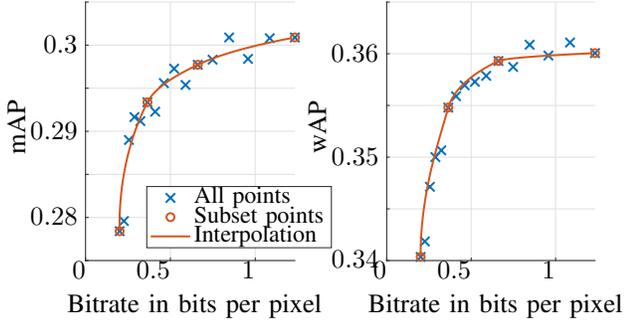}
\caption{Subset points (o's), all points (x'es) and interpolated curves (Akima interpolation) for the mAP-rate performance (left) and wAP-rate performance (right) of Yolov5 (high quality).   } 
\label{fig:interpAccEx3}
\end{figure}

We found that in certain cases, the performance metrics show a noisy behavior leading to  non-monotonous values. An example is shown in Fig.~\ref{fig:interpAccEx3} on the left, where the mAP is prone to significant noise. 
In this case, conclusions from interpolated curves are hard to draw, in particular because the subset points show a reasonable (i.e., monotonous) behavior. Here, the reason for the high variability is the averaging over classes as performed for mAP calculations. The mAP is averaged over a set of object classes, where some object classes only contain very few instances (e.g., the images in the set include only a couple of trains, such that the detection accuracy for trains cannot be evaluated properly). A reasonable solution is to use the wAP value as proposed in \cite{Fischer21}, where the average is not calculated over the classes, but over the number of object instances. As a consequence, the noise is reduced significantly. This is reflected by a lower variability in supporting points (see Fig.~\ref{fig:interpAccEx3} on the right) and the mean RIE values ($4.7\%$ vs. $8.5\%$ for wAP-bitrate and mAP-bitrate, respectively, in low-QP VCM, MaskRCNN, Appendix E). 

Furthermore, we observed that if the independent variable (the quality metric) of the supporting points is not monotonous, the interpolation and thus the BD value cannot be calculated at all. Hence, a check for monotony for the independent metrics must always be performed. If the dependent variable is not monotonous, interpolation does not lead to algorithmic problems because in PCHIP and Akima interpolation, such points will be interpreted as maxima or minima of the interpolated curve (see Section~\ref{sec:bjonteCalc}). The latter case was sometimes observed for complexity metrics such as processing time or energy. Note that non-monotonous behavior can point to errors or indicate \CCH{inadequately} designed testing procedures. 

\subsection{Relative Curve Difference (RCD)}
\label{sec:RCD}

\begin{figure}
\psfrag{011}[tc][tc]{ Relative bitrate difference (\%)}%
\psfrag{012}[bc][bc]{ PSNR in dB}%
\psfrag{000}[r][ct]{ $-0.3$}%
\psfrag{001}[r][ct]{ $-0.2$}%
\psfrag{002}[r][ct]{ $-0.1$}%
\psfrag{003}[r][ct]{ $0$}%
\psfrag{004}[r][ct]{ $0.1$}%
\psfrag{005}[r][ct]{ $0.2$}%
\psfrag{006}[rc][rc]{ $34$}%
\psfrag{007}[rc][rc]{ $36$}%
\psfrag{008}[rc][rc]{ $38$}%
\psfrag{009}[rc][rc]{ $40$}%
\psfrag{010}[rc][rc]{ $42$}%
\psfrag{PCHIP, supportttt}[l][l]{\small RCD}%
\psfrag{PCHIP all}[l][l]{\small $\BD=-0.08\%$}%
\psfrag{Akima support}[l][l]{\small Supp. points B}%
\psfrag{Akima all}[l][l]{\small Supp. points A}%
\includegraphics[width=0.5\textwidth]{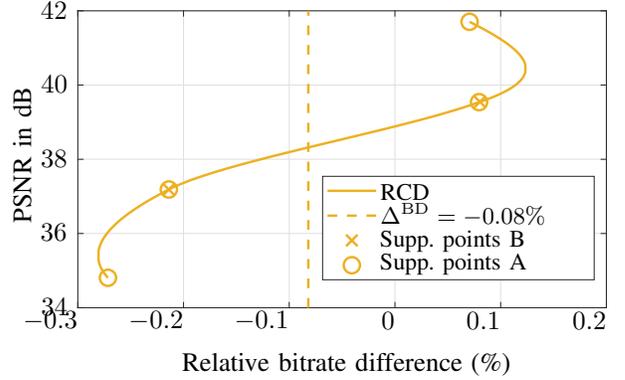}
\caption{Relative curve difference (RCD) $\Delta P_\mathrm{dep}(P_\mathrm{ind})$ between two RD curves as calculated by Eq.~\eqref{eq:rdDiff}, Akima interpolation. In this example, the vertical axis $P_\mathrm{ind}$ is the PSNR. The performance points are taken from the RAT data set, where the tool Luma Mapping with Chroma Scaling (LMCS) \cite{Lu20} is switched on (encoder A) and off (encoder B). The sequence is BQMall, the encoder is VVenC with the `slower' preset. The solid line represents the interpolated curve difference using four subset points. The vertical dashed line denotes the $\BD$-rate value. The markers denote the positions of the supporting points inside the integration bounds, where the `x'-markers denote the reference points (encoder B) and the `o'-markers the test points (encoder A). }
\label{fig:BDdiff}
\end{figure}

\begin{figure}
\psfrag{012}[t][tc]{  Relative bitrate difference (\%)}%
\psfrag{013}[bc][bc]{ PSNR in dB}%
\psfrag{000}[r][ct]{ $0$}%
\psfrag{001}[r][ct]{ $5$}%
\psfrag{002}[r][ct]{ $10$}%
\psfrag{003}[r][ct]{ $15$}%
\psfrag{004}[r][ct]{ $20$}%
\psfrag{005}[r][ct]{ $25$}%
\psfrag{006}[rc][rc]{ $34$}%
\psfrag{007}[rc][rc]{ $35$}%
\psfrag{008}[rc][rc]{ $36$}%
\psfrag{009}[rc][rc]{ $37$}%
\psfrag{010}[rc][rc]{ $38$}%
\psfrag{011}[rc][rc]{ $39$}%
\psfrag{Akima support}[l][l]{\small Subset RCD}%
\psfrag{PCHIP all}[l][l]{\small All RCD}%
\psfrag{BD value}[l][l]{\small $\BD$}%
\psfrag{Support Pointsss}[l][l]{\small Support points (A)}%
\includegraphics[width=.49\textwidth]{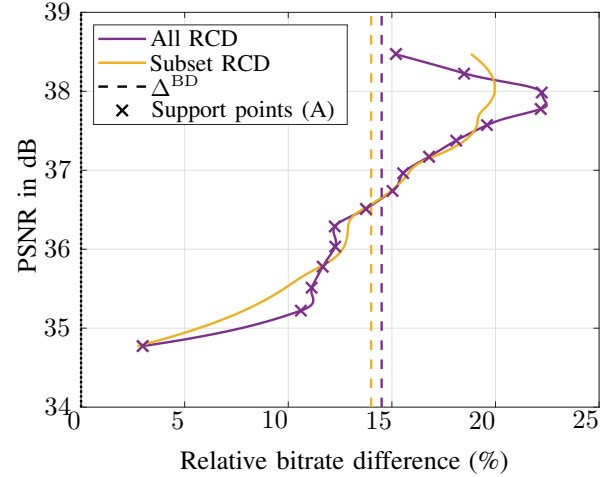}
\caption{RCDs of the BQTerrace sequence for VTM encoding with ALF on and off using the PSNR and the bitrate as performance metrics. The dashed vertical lines indicate the BD-value for the subset points and all points. }
\label{fig:BDdiffAll}
\end{figure}

As we have seen in Section~\ref{sec:interAcc}, none of the tested interpolation methods can perfectly reconstruct intermediate performance points. Consequently, this problem could lead to inaccurate $\BD$ values. To find out about the impact of the interpolation error on the $\BD$, we first introduce a new tool that helps in understanding and interpreting $\BD$ values. The tool is an intermediate result of the $\BD$ calculations which uses the two interpolated curves to calculate the difference curve. This helps to visualize the relative performance depending on the independent variable. The tool, the relative curve difference (RCD), corresponds to Eq.~\eqref{eq:bdIntegration} without the integration and is calculated as 
\begin{equation}
\Delta P_\mathrm{dep}(P_\mathrm{ind})\, [\%] = 100 \cdot \left( \frac{10^{\hat p_\mathrm{dep}(\mathrm{A},P_\mathrm{ind})}}{10^{\hat p_\mathrm{dep}(\mathrm{B},P_\mathrm{ind})}}-1\right). 
\label{eq:rdDiff}
\end{equation} 
\CCH{Note that it can be calculated for any number of support points, as shown in Figs.~\ref{fig:BDdiff} and \ref{fig:BDdiffAll}.  We discuss three examples in which the RCD provides valuable information. The first example} is illustrated in Fig.~\ref{fig:BDdiff}. 

The figure shows the interpolated curve difference between two encoder configurations A and B. 
The solid curve is the difference when using the four subset RD points ('x`-markers and 'o`-markers) for interpolation. 


The plot shows the compression performance differences depending on the independent variable, e.g., the visual quality. While the $\BD$ represents mean savings, the RCD represents the actual savings at a fixed quality. In this cherry-picked example, we can see that the curve difference changes its sign at $P_\mathrm{ind}\approx 39\,$dB, which corresponds to intersecting RD curves. This is not reflected by the $\BD=-0.08\%$, which indicates that encoder A has a higher compression performance than encoder B. Hence, the RCD is useful for compression tool developers to get information on certain quality ranges in which a coding tool has a higher performance.

%

\begin{figure}
\psfrag{014}[tc][tc]{ Relative bitrate difference (\%)}%
\psfrag{015}[bc][bc]{$-10 \cdot \log_\mathrm{10}\left(1-\mathrm{SSIM}\right)$ in dB }%
\psfrag{Points ABCDEF}[l][l]{Supp. points A}%
\psfrag{Points B}[l][l]{Supp. points B}%
\psfrag{RCD}[l][l]{RCD}%
\psfrag{BD-Rate}[l][l]{$\Delta^\mathrm{BD}$}%
\psfrag{000}[r][ct]{ $ -10$}%
\psfrag{001}[r][ct]{ $ -8$}%
\psfrag{002}[r][ct]{ $ -6$}%
\psfrag{003}[r][ct]{ $ -4$}%
\psfrag{004}[r][ct]{ $ -2$}%
\psfrag{005}[r][ct]{ $0$}%
\psfrag{006}[rc][rc]{ $10$}%
\psfrag{007}[rc][rc]{ $12$}%
\psfrag{008}[rc][rc]{ $14$}%
\psfrag{009}[rc][rc]{ $16$}%
\psfrag{010}[rc][rc]{ $18$}%
\psfrag{011}[rc][rc]{ $20$}%
\psfrag{012}[rc][rc]{ $22$}%
\psfrag{013}[rc][rc]{ $24$}%
\includegraphics[width=.5\textwidth]{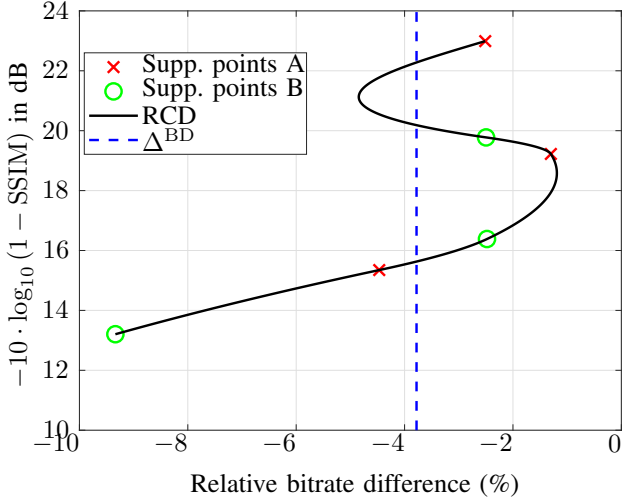}
\caption{\CCH{Example of an RCD curve with an overshoot. The example is taken from the ENC dataset, the PM is log SSIM vs. bitrate. }}
\label{fig:RCD_overshoot}
\end{figure}

\CCH{The second example is illustrated in Fig.~\ref{fig:RCD_overshoot}. In this figure, we can see the RCD for SSIM in log-domain vs. the bitrate (ENC dataset). The red and the green markers correspond to the vertical position of the support points. We can see that the interpolated curve difference shows an overshoot between \CCHT{log SSIM} values of $20$ and $23$. As in the vicinity of this region,  the relative difference is rather small (around $-2.5\%$), we would expect similar values here. The overshoot indicates that the interpolation could be inaccurate\footnote{Note that such an overshoot does not indicate that the interpolated curves also show an overshoot, only the interpolated curve difference does so.}.    }

\CCH{As a third example, we show the impact of disabling ALF in VTM encoding in Fig.~\ref{fig:BDdiffAll}.  In this plot, we show two RCD curves: first, the RCD for the subset points (yellow), and second, the RCD for all points (purple). We can observe two issues: First, we find that the range of bitrate differences is large ($5\%$ up to $20\%$). The plot indicates that the ALF is most effective at high qualities and much less effective at lower qualities. This information is not reflected by the constant $\BD$ of $14\%$. Second, the RCD for all supporting points shows that subsampling of QP values can lead to missed data points. In this particular case, the $\BD$ for all points ($14.5\%$) is higher than for the subset points ($14\%$), because the large relative differences above $20\%$ (purple markers on the right) were not detected using the subset points. }
In the following subsection, using RCD curves, we will motivate and evaluate another error metric to assess the reliability of $\BD$ values. 

\subsection{BD Accuracy}
\label{sec:BDAcc}

The preceding subsections have shown that current interpolation methods for $\BD$ calculations show some deficiencies. Consequently, the question arises whether these deficiencies also lead to inaccuracies in the final $\BD$, which compares two interpolated curves. For example in codec standardizations, decisions to adopt or reject a new coding tool are usually based on a RD comparison between a reference encoder with and without a certain coding tool. In many cases, the reported savings yield small values close to $1\%$ \cite{Lu20,DeLuxanHernandez19,Kraenzler21}, which is in the same range as the mean interpolation errors reported before. Hence, in this subsection, we will discuss whether and in which cases the aforementioned interpolation errors can lead to misleading interpretations. 

To this end, we propose to evaluate the accuracy of the classic calculation using a subset of supporting points $\BDsub$ by comparing with the value obtained using all points $\BDall$ as visualized in Fig.~\ref{fig:BDdiffAll}.  In the figure, we can see the RCD for the subset points \CCH{and for all points  (yellow and purple, respectively) and their corresponding $\BD$ values. } The purple RCD is the most accurate representation of the bitrate difference between the two encoders A and B that we can get, because the sampling of the supporting points is as dense as possible. As such, the corresponding $\BDall$ is more accurate than the subset $\BDsub$. Thus, we propose to use the difference between these two $\BD$ values as an error metric. 

For one sequence $s$ and a certain configuration $\kappa$, we call this error metric the ``subset error'' and calculate it by 
\begin{equation}
 e^\mathrm{sub}_{s,\kappa}  =  \Delta^\mathrm{BD}_{\mathrm{sub},s,\kappa} -  \Delta^\mathrm{BD}_{\mathrm{all},s,\kappa}. 
\label{eq:subsErr}
\end{equation}
We obtain the mean subset error by averaging over all tested sequences $s\in \mathcal{S}$ and configurations $\kappa\in \mathcal{C}$, excluding the reference configuration needed for $\BD$ calculations, as 
\begin{equation}
\bar e^\mathrm{sub}  = \frac{1}{|\mathcal{S}||\mathcal{C}|}\sum_{s\in \mathcal{S}}\sum_{\kappa\in\mathcal{C}} \left| e^\mathrm{sub}_{s,\kappa}\right|
\label{eq:BD4vsAllError}
\end{equation}
\CCH{and additionally report the standard deviation of the error }
\begin{equation}
e_\sigma^\mathrm{sub}  = \sqrt{\frac{1}{|\mathcal{S}||\mathcal{C}|}\sum_{s\in \mathcal{S}}\sum_{\kappa\in\mathcal{C}} \left(e^\mathrm{sub}_{s,\kappa} - e_\mathrm{mean}^\mathrm{sub} \right)^2}, 
\label{eq:BD4vsAllError_max}
\end{equation}
\CCH{where $e_\mathrm{mean}^\mathrm{sub}$ is the mean subset error.} 

\CCH{In the following, first, we compare PCHIP with Akima. Afterwards, we investigate the impact of the number of supporting points. Finally, we report and discuss observed subset errors for different datasets. }

\subsubsection{PCHIP vs. Akima Interpolation}
\begin{figure}
\psfrag{006}[bc][bc]{ Aboslute subset error $|e^\mathrm{sub}|$}%
\psfrag{000}[r][ct]{ Akima}%
\psfrag{001}[r][ct]{ PCHIP}%
\psfrag{002}[rc][rc]{ $0\%$}%
\psfrag{003}[rc][rc]{ $1\%$}%
\psfrag{004}[rc][rc]{ $2\%$}%
\psfrag{005}[rc][rc]{ $3\%$}%
\includegraphics[width=.5\textwidth]{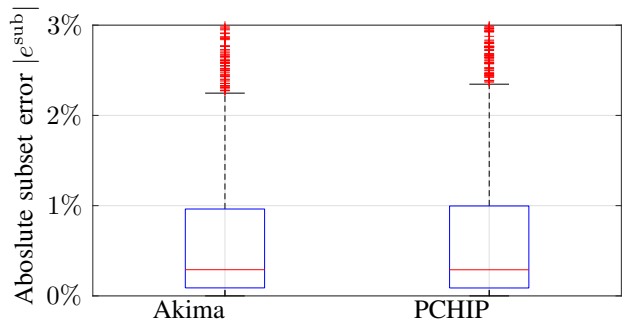}
\caption{Box-whisker plot on all subset errors for Akima interpolation (left) and PCHIP (right). The vertical axis denotes the absolute subset errors w.r.t. $\BDall$ interpolated by Akima interpolation. }
\label{fig:subsetErrorBoxes}
\end{figure}

\CCH{We compare the impact of the interpolation method as follows. First, we calculate $\BDsub$ values for Akima interpolation and PCHIP. The subset error $e^\mathrm{sub}$ is then obtained using the $\BDall$ calculated with Akima interpolation, because it returns more accurate interpolated curves (see Section~\ref{sec:interAcc}). The absolute subset errors for all instances from all datasets are collected and a statistical analysis using box-whisker plots is illustrated in Fig.~\ref{fig:subsetErrorBoxes}. }

\CCH{The plot indicates that the distribution of subset errors is very similar for PCHIP and Akima, because the median, the box boundaries, and the whiskers are located at similar vertical positions. To have a further indication, we perform a two-sample $t$-test \cite{Bendat71} to check whether the two distributions (Akima and PCHIP) have the same mean value, which is the null-hypothesis. The test rejects the alternative hypothesis that the mean value is different at a significance level of $5\%$ with a $p$-value of $73\%$. Hence, we conclude that actual $\BD$ values calculated by PCHIP are not significantly different from $\BD$ values calculated by Akima interpolation, such that both interpolation methods are a valid choice to calculate the $\BD$.}

\subsubsection{Number of Supporting Points}
To reduce the subset error, it was recommended to \CCH{increase the number of} supporting points  (e.g., in the AOM common test conditions \cite{AOM_CTC21} using six supporting points \CCH{instead of four used in JVET \cite{JVET_T2010})}. Hence, we perform a dedicated analysis with different numbers $I$ of supporting points using the RAT data set. For this, we split the standard QP range into $I-1$ intervals and choose a corresponding number of input QP values. Using the standard QP range (in our case QP$\,\in\{22,23,...,37\}$), we often cannot split the original interval into equidistant intervals in the QP domain. As the full range is $\Delta\mathrm{QP}=15$, equidistant points are only possible for $I-1=1$, $I-1=3$, and $I-1=5$ intervals. For other interval numbers, we round the fractional QPs to the nearest integer value. The corresponding tested numbers of supporting points $I$ and the supporting QPs are listed in Table~\ref{tab:nQPs}. 
\begin{table}[t]
\renewcommand{\arraystretch}{1.3}
\caption{Tested numbers of supporting points $I$ and selection of QP values for evaluating the subset error $\bar e^\mathrm{sub}$.   }
\label{tab:nQPs}
\vspace{-0.3cm}
\begin{center}
\begin{tabular}{l|l}
\hline
$I$ & QPs\\
\hline
2 & $\{22,37\}$ \\ 
3 & $\{22, 30, 37\}$ \\ 
4 & $\{22, 27, 32, 37\}$ \\ 
5 & $\{22, 26, 30, 34, 37\}$ \\ 
6 & $\{22, 25, 28, 31, 34, 37\}$ \\ 
7 & $\{22, 24, 27, 29, 32, 34, 37\}$ \\ 
8 & $\{22, 24, 26, 28, 30, 32, 34, 37\}$ \\ 
9 & $\{22, 24, 26, 28, 30, 32, 34, 36, 37\}$ \\ 
 \hline
 \end{tabular}
\end{center}
\end{table}
We chose a maximum of $I=9$ supporting points, which corresponds to using every second QP as a supporting point. Note that $I=2$ \CCHT{leads to a linear and $I=3$ leads to quadratic polynomial instead of the cubic polynomial in Eq.~\eqref{eq:BDpoly}}.

\begin{figure}
\psfrag{011}[tc][tc]{ Number of supporting points $I$}%
\psfrag{012}[bc][bc]{ $\bar e^\mathrm{sub}$ in \%}%
\psfrag{000}[r][ct]{ $2$}%
\psfrag{001}[r][ct]{ $3$}%
\psfrag{002}[r][ct]{ $4$}%
\psfrag{003}[r][ct]{ $5$}%
\psfrag{004}[r][ct]{ $6$}%
\psfrag{005}[r][ct]{ $7$}%
\psfrag{006}[r][ct]{ $8$}%
\psfrag{007}[r][ct]{ $9$}%
\psfrag{008}[rc][rc]{ $0$}%
\psfrag{009}[rc][rc]{ $0.5$}%
\psfrag{010}[rc][rc]{ $1$}%
\psfrag{data1aaaaaaaaaaaaa}[l][l]{\small{PSNR-bitrate}}%
\psfrag{data2}[l][l]{\small{PSNR-dec. energy}}%
\psfrag{data3}[l][l]{\small{log VMAF-bitrate}}%
\includegraphics[width=.49\textwidth]{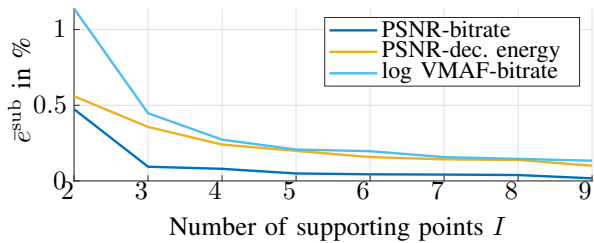}
\caption{Mean subset errors depending on the number of supporting points $I$ (Akima interpolation). }
\label{fig:errOverSubset}
\end{figure}

For each subset, we then calculate the mean subset error and plot the result in Fig.~\ref{fig:errOverSubset}. 
The dark blue line represents the mean subset error for PSNR-bitrate. In BD calculations as in the CTC \cite{JVET_T2010} with $I=4$, we can observe a mean error of $0.08\%$. Interestingly, the value is only slightly higher with $I=3$ ($0.09\%$). Using two supporting points (linear interpolation), we still reach a mean estimation error of $0.47\%$, which can be sufficiently accurate when large BD differences shall be reported. 

Using $I=6$ supporting points leads to an \CCHT{improved accuracy with an error of $\bar e^\mathrm{sub}=0.044\%$, which is almost halved with respect to $I=4$.  This }can be beneficial when curve differences are small. At higher numbers ($I\le8$), the error barely changes. The error drops again if $I=9$ because in this case, at least every second QP is used as a supporting point (cf. Table~\ref{tab:nQPs}). Here, we reach a mean error of $0.02\%$. 


We show the curves for two more performance metrics (PSNR-decoding energy in yellow and log VMAF-bitrate in bright blue). They show a similar trend as the PSNR-bitrate errors, but at a higher error level. Summarizing, we find that two or three supporting points are sufficient to generate rough performance comparisons, which is beneficial if processing times shall be reduced (encoding and evaluation for two instead of four QPs). 

\subsubsection{Subset Errors for Datasets}

\CCH{Due to the high amount of data, we only discuss a part of all subset errors in this section. The remaining subset errors are available in the supplemental material, \CCHT{Appendix E}. First, we list values for four supporting points and for six supporting points in Table~\ref{tab:RAT_BDerr} (RAT dataset, all PM pairs).} 

\begin{table*}[t]
\renewcommand{\arraystretch}{1.3}
\caption{Subset error values $\bar e^\mathrm{sub}$ and $e_\sigma^\mathrm{sub}$ for data set RAT with Akima interpolation for four supporting points and six supporting points.   }
\label{tab:RAT_BDerr}
\vspace{-0.3cm}
\begin{center}
\resizebox{\textwidth}{!}{ 
\begin{tabular}{l|r|r|r|r|r|r|r|r|r|r}
\hline
& \multicolumn{2}{c|}{PSNR-bitrate} & \multicolumn{2}{c|}{PSNR-Time} & \multicolumn{2}{c|}{PSNR-Dec. Energy} 
& \multicolumn{2}{c|}{log VMAF-Rate} & \multicolumn{2}{c}{log VMAF-Energy}\\
&  $\bar e^\mathrm{sub}$ & $e_\sigma^\mathrm{sub}$ & $\bar e^\mathrm{sub}$ & $e_\sigma^\mathrm{sub}$ & $\bar e^\mathrm{sub}$ & $e_\sigma^\mathrm{sub}$ & $\bar e^\mathrm{sub}$ & $e_\sigma^\mathrm{sub}$ & $\bar e^\mathrm{sub}$ & $e_\sigma^\mathrm{sub}$ 
\\ \hline
$4$ supporting points & \small $0.080\%$ & \small $0.148\%$  & \small $0.182\%$ & \small $0.290\%$  & \small $0.240\%$ & \small $0.329\%$  & \small $0.273\%$ & \small $0.478\%$  & \small $0.284\%$ & \small $0.391\%$ \\ 
$6$ supporting points & \small $0.044\%$ & \small $0.085\%$  & \small $0.107\%$ & \small $0.176\%$  & \small $0.158\%$ & \small $0.215\%$  & \small $0.197\%$ & \small $0.350\%$  & \small $0.174\%$ & \small $0.225\%$ \\ 
 \hline
 \end{tabular}}
\end{center}
\end{table*}

The table shows that all mean subset errors $\bar e^\mathrm{sub}$ are significantly lower than interpolation errors $e^\mathrm{RIE}$ illustrated  in Figure~\ref{fig:boxplotInterpError}. We observed this behavior for almost all tested cases. \CCH{Considering the standard deviation of the errors $e_\sigma^\mathrm{sub}$, we can see that they are significantly higher than the mean subset errors, which proves the high variability of the subset errors. }
\CCH{Still, as both the mean subset error and the standard deviation are significantly smaller than the mean interpolation error (e.g., $0.4\%$ for PSNR-bitrate, RAT dataset), }we can draw the conclusion that by comparing two curves, a part of the interpolation error is averaged. \CCH{However,} concerning PSNR-bitrate comparisons in the RAT data set conditions, a $\BD$ difference below $0.15\%$ (cf. $e_\sigma^\mathrm{sub}=0.148\%$ for four supporting points) could be misleading. Furthermore, we can see that subset errors are lowest for classic PSNR-bitrate considerations. For example, using VMAF instead of the PSNR, the mean subset error increases to more than $0.25\%$. 

\begin{figure}
\psfrag{010}[bc][bc]{ Subset errors}%
\psfrag{000}[rt][cb]{ \rotatebox{30}{{PSNR}-bitrate}}%
\psfrag{001}[rt][cb]{ \rotatebox{30}{{PSNR}-time}}%
\psfrag{002}[rt][cb]{ \rotatebox{30}{{PSNR}-dec. energy}}%
\psfrag{003}[rt][cb]{ \rotatebox{30}{{log VMAF}-bitrate}}%
\psfrag{004}[rt][cb]{ \rotatebox{30}{log VMAF-dec. energy}}%
\psfrag{005}[rc][rc]{ $0$}%
\psfrag{006}[rc][rc]{ $0.5$}%
\psfrag{007}[rc][rc]{ $1$}%
\psfrag{008}[rc][rc]{ $1.5$}%
\psfrag{009}[rc][rc]{ $2$}%
\includegraphics[width=.5\textwidth]{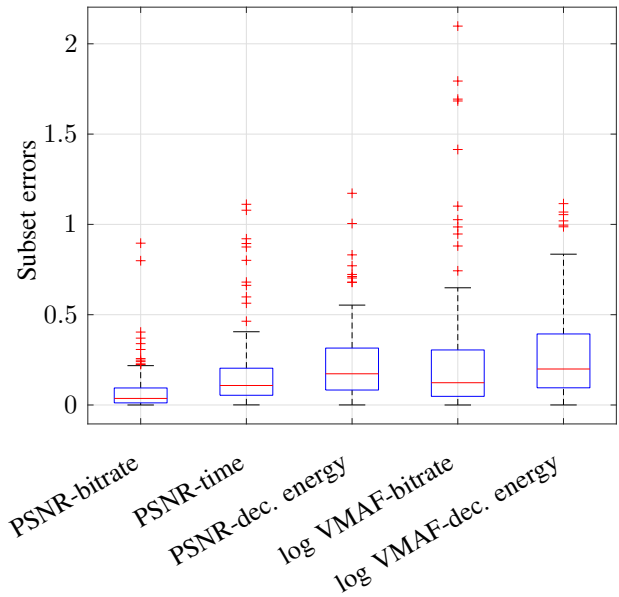}
\vspace{1.2cm}
\caption{\CCH{Distribution of subset errors (four supporting points) for all PM pairs of the RAT dataset. }}
\label{fig:RATdistr}
\end{figure}

\CCH{Statistical distributions of the corresponding subset errors are illustrated in Fig.~\ref{fig:RATdistr}. We can see that apart from some outliers, all subset errors are smaller than $1\%$.  }
\CCH{We further discuss subset errors for selected PM pairs of all datasets in Table~\ref{tab:datasetSubs}. Remaining values are listed in Appendix E. }
\begin{table}
\caption{\CCH{Selected subset errors for all datasets. The standard deviation reported for VCM is zero because  it consists of a single sample. } }
\label{tab:datasetSubs}
\centering
\begin{tabular}{l|c|r|r}
\hline
Dataset & PM pair & $\bar e^\mathrm{sub}$ & $e_\sigma^\mathrm{sub}$\\
\hline
LDT & PSNR-bitrate & $0.073\%$ & $0.124\%$ \\
CC & PSNR-bitrate & $0.462\%$ & $0.323\%$ \\
CC & VMAF-bitrate & $1.210\%$ & $1.154\%$\\
CC & log VMAF-bitrate & $1.164\%$ & $1.612\%$\\
ENC & PSNR-bitrate & $0.340\%$ & $0.450\%$ \\
ENC & SSIM-bitrate & $0.959\%$ & $0.702\%$ \\
ENC & log SSIM-bitrate & $0.357\%$ & $0.448\%$ \\
ENC & SSIM-enc. energy & $8.598\%$ & $18.252\%$ \\
ENC & log SSIM-enc. energy & $5.370\%$ & $11.260\%$ \\
AOM & PSNR-bitrate & $2.267\%$ & $3.848\%$ \\
NNV & PSNR-bitrate & $0.484\%$ & $0.959\%$ \\
SCC & PSNR-RGB-bitrate & $0.125\%$ & $0.101\%$ \\
VCM (high QP) & mIoU-bitrate &  $1.763\%$ & $0.000\%$ \\
VCM (low QP) & wAP-bitrate FasterRCNN & $4.461\%$ & $0.000\%$  \\
360 & Y-E2EPSNR-NN-filesize & $0.311\%$ & $0.338\%$ \\
PCA & Y-PSNR-bitrate & $0.646\%$ & $0.620\%$ \\
PCG & d2-PSNR-bitrate & $5.360\%$ & $3.400\%$ \\
\hline
\end{tabular}
\end{table}
First, the table indicates that very low subset errors (below $0.1\%$) are only achieved for tool tests (RAT and LDT dataset). When comparing different codecs (e.g., CC, NNV) or different encoder configurations (e.g., ENC, AOM), mean subset errors are usually significantly higher ($0.1\%$ up to $2.3\%$). Note that very high errors (e.g., $2.3\%$ for AOM) are observed when the $\BD$ values are also very high. In the case of AOM \CCHT{with PMs PSNR and bitrate}, $\BD$ values greater than $100\%$ \CCHT{occur such that in relation to the observed maximum $\BD$ of $805\%$, the corresponding subset error of $16\%$ is} relatively small. 

\CCH{Furthermore, we can observe large errors for VCM, PCG, and ENC SSIM-encoding energy  ($4.6\%$, $5.4\%$, and $8.6\%$, respectively), which indicates that BD calculations are less accurate in these uncommon testing conditions. For corresponding analyses, we strongly recommend to report more data than just the $\BD$, for example, RCD curves.  }

\CCH{Finally, we check subset errors for saturating metrics such as VMAF and SSIM. In Subsection~\ref{sec:satMetrics}, it was claimed that the interpolation is more accurate using a logarithmic representation of VMAF and SSIM. To show the impact on final $\BD$ values, we present subset errors for the PM pairs VMAF-bitrate, SSIM-bitrate, and SSIM-encoding energy in Table~\ref{tab:datasetSubs}. We can see that in the logarithmic domain, the subset error decreases slightly for VMAF-bitrate ($1.210\%$ to $1.164\%$). In case of SSIM, the error decreases substantially from $0.96\%$ to $0.36\%$ and from $8.6\%$ to $5.4\%$ (SSIM-bitrate and SSIM-encoding energy, respectively). Hence, we conclude that in general, the logarithmic representation is beneficial for $\BD$ calculations. }

\subsubsection{Performance Metric Ranges}
\begin{figure}
\psfrag{007}[c][c]{ Bitrate in kbps}%
\psfrag{008}[bc][bc]{ PSNR in dB}%
\psfrag{000}[r][ct]{$1$}%
\psfrag{001}[r][ct]{$ 10$}%
\psfrag{002}[rc][rc]{ $34$}%
\psfrag{003}[rc][rc]{ $36$}%
\psfrag{004}[rc][rc]{ $38$}%
\psfrag{005}[rc][rc]{ $40$}%
\psfrag{006}[rc][rc]{ $42$}%
\psfrag{data1aaaaaaaa}[l][l]{\small Interp. A}%
\psfrag{data2}[l][l]{\small Interp. B }%
\psfrag{data3}[l][l]{\small All points}%
\psfrag{data4}[l][l]{\small Subset points}%
\includegraphics[width=.49\textwidth]{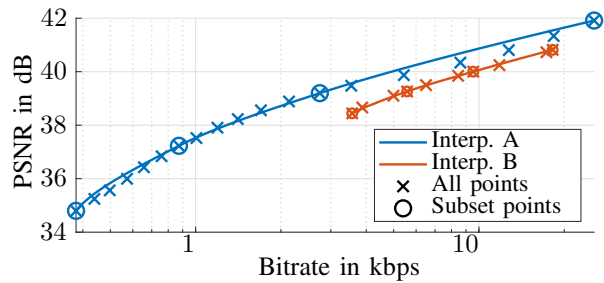}
\caption{Performance curves of the videoSRC15 sequence for two different encoders using the PSNR and the bitrate as PMs. The encoders are DMC-32 (A) and HM (B). }
\label{fig:badIndRange}
\end{figure}

In this subsection, we discuss cases in which the range of the PMs differ significantly for the two encoders A and B. For the independent PM, this can be observed in the literature when different codecs or compression technologies are compared (e.g., \cite{Li2022,Sheng22,Minallah15}). We show an example from the NNV dataset and plot the corresponding supporting points and interpolated curves in Fig.~\ref{fig:badIndRange}. 
We can see that the blue curve, which interpolates HM encoding, shows an inaccurate interpolation at a PSNR of roughly $40\,$dB. As a consequence, the $\BD$ using the subset points 
yields approximately $-49\%$ while using all available supporting points leads to a more realistic value of $\BDall\approx -43\%$. 

The reason for this large deviation is that in the overlapping PSNR range, we can only find a single supporting point from the HM data (blue circle marker at $39\,$dB), leading to a large interpolation error. 
We conclude that the supporting points of both encoders should be chosen such that the intersection of the independent variable is as high as possible. \CCH{To facilitate controlling this condition, we implemented an algorithm calculating the intersection over union (IoU) of the two ranges, which can be used to validate the choice of the supporting points \cite{BD_LMS}.}

The second case corresponds to large differences in the dependent variable \CCHT{(e.g., the rate or the energy),} which we already discussed in the previous subsection. In this case, $\BD$ values can be very large (AOM dataset\CCHT{, PMs PSNR and bitrate, with a $\BD$ value up to $800\%$}), which also leads to large subset errors. As in many cases, a large $\BD$ value is desired, we do not consider this effect problematic. However, we conclude that the \CCHT{quality metric should always be defined as the independent variable}, because the range of desired qualities \CCHT{is stable}, while \CCHT{values of }other metrics such as the encoding time or the bitrate can \CCHT{be located in disjoint ranges.}

For example \CCHT{in encoding}, fast presets lead to encoding times which are smaller by several orders of magnitude than slow presets. Hence, on the time axis, the performance curves of a slow and a fast preset may not overlap, even if a high \CCHT{quality range} is covered. Consequently, a BD metric using the time as the independent variable cannot be evaluated because the integration interval is void. As such, it is recommended to not use the BD-PSNR metric. 

%

As a conclusion to this section, we claim that the \CCHT{mean and the standard deviation of the subset error} can be used as a statistical metric indicating the expected error of $\BD$ values. As such, the data on subset errors provided in \CCHT{Appendix E} can be used to check whether a practically calculated $\BD$ value represents a reliable tendency. 

\section{Recommendations}
\label{sec:recs}
This section summarizes all recommendations derived from our study: 
\begin{itemize}
\item Akima interpolation \CCH{returns more accurate interpolated curves. $\BD$ values are similar for both Akima and PCHIP. }
\item The interpolation method should be reported when providing $\BD$ values, in particular when using legacy interpolation methods \CCH{such as CSI}. 
\item For VMAF and SSIM, the logarithmic value as calculated by Eq.~\eqref{eq:logSSIM} and \eqref{eq:logVMAF} should be used, respectively.
\item Monotony of the independent variable must be assured.  
\item Two supporting points are sufficient to obtain rough performance estimates. 
\item The range of values of the independent variable (e.g., the quality metric) should have a high intersection for the two encoder configurations. \CCH{Intersection over union values should be checked. }
\item \CCH{If the mean $\BD$ value is smaller than the mean subset error $\bar e^\mathrm{sub}$ of the corresponding dataset and performance metrics, conclusions on the actual compression performance should be drawn carefully. E.g., } the RCD from Eq.~\eqref{eq:rdDiff} should be investigated and the generation of further performance points should be considered. 
\end{itemize}

For practical use, we provide implementations for Python and Matlab, which return the $\BD$, \CCH{the IoU, and plot the RCD using PCHIP and }Akima interpolation \cite{BD_LMS}. The implementation supports any number of supporting points greater than two, where $I=2$ reverts to linear and $I=3$ to square interpolation. 
 The implementations can  be used under the BSD-3-Clause license.

\section{Conclusion}
\label{sec:concl}
This paper has shown an explanation and analysis on the Bj{\o}ntegaard-Delta calculations. We summarized major literature on this metric and collected practical use cases. Furthermore, we provided an understandable mathematical formulation. We performed an analysis on a large database and came up with a set of recommendations on the correct and reliable usage of the metric, which includes the use of Akima interpolation and the use of the logarithmic domain for SSIM and VMAF. In future work, we will analyze the BD calculus in further applications including subjective quality scores.

\ifCLASSOPTIONcaptionsoff
  \newpage
\fi

\bibliographystyle{IEEEbib}
\bibliography{literature}

%
%
%




\clearpage

\appendix

\section*{A. Mathematical Definition of Interpolation Methods}
In this appendix, we describe the three interpolation methods CSI, PCHIP, and Akima mathematically. The supporting points are indicated by $\{P_\mathrm{dep}(i),P_\mathrm{ind}(i)\}, \, i=1,...,I$, where for simplicity, we drop the index $k$, which usually represents one of two encoders or encoder configurations. Correspondingly, the interpolated curve is denoted by $\hat p_\mathrm{dep}(P_\mathrm{ind})$. It is defined as a third-order piecewise polynomial function (see Eq.~(3)), where $I-1$ pieces must be determined when $I$ supporting points are available. This results is $M=4\cdot(I-1)$ coefficients. 

All interpolation methods share the property that the interpolated curve shall pass the supporting points. As a consequence, the first set of conditions for the $i$-th piece of the piecewise polynomial is given as 
\begin{align}
\hat p_{\mathrm{dep},i}(P_\mathrm{ind}(i)) & = \log_{10}\left(P_\mathrm{dep}(i)\right), \label{eq:interpoConstBase}\\
\hat p_{\mathrm{dep},i}(P_\mathrm{ind}(i+1)) & = \log_{10}\left(P_\mathrm{dep}(i+1)\right).   \nonumber
\end{align}
This results in two conditions per piece or $2(I-1)$ conditions, such that $2I-2$ more conditions are needed. 

Furthermore, for all internal supporting points, it is desired to have smooth transitions. This is achieved by defining a constant derivative from both sides of the supporting points given by 
\begin{equation}
\hat p'_{\mathrm{dep},i}(P_\mathrm{ind}(i+1)) = \hat p'_{\mathrm{dep},i+1}(P_\mathrm{ind}(i+1)), \label{eq:slopeConstr}
\end{equation}
which leads to $I-2$ additional conditions. 
The remaining $I$ conditions depend on the interpolation method and are described in the subsequent subsections.

\subsection*{1) Cubic Spline Interpolation}
\label{secsec:CSI}
Cubic spline interpolation (CSI) is a special case of spline interpolation with a third-order polynomial as a basis. The remaining $I$ conditions are obtained by setting a constant curvature at all internal supporting points given by  \cite{Young17} 
\begin{equation}
\hat p''_{\mathrm{dep},i}(P_\mathrm{ind}(i+1)) = \hat p''_{\mathrm{dep},i+1}(P_\mathrm{ind}(i+1)), \label{eq:curvConstr}
\end{equation}
which leads to $I-2$ additional conditions. 
For the remaining two conditions, in standardization it was proposed to use the ``not-a-knot'' condition \cite{Bjonte01}, in which third order derivatives at the supporting points next to the boundaries are set constant as
\begin{align}
\hat p'''_{\mathrm{dep},1}(P_\mathrm{ind}(1)) & = \hat p'''_{\mathrm{dep},2}(P_\mathrm{ind}(1))\label{eq:notKnotConstr1}\\
\hat p'''_{\mathrm{dep},I-2}(P_\mathrm{ind}(I-1)) & = \hat p'''_{\mathrm{dep},I-1}(P_\mathrm{ind}(I-1)). 
 \label{eq:notKnotConstr2}
\end{align}
Thus, all coefficients can be determined. 

It is worth mentioning that in the case of four supporting points, the resulting third-order polynomial is not piecewise, but a single polynomial valid for each interval of the interpolated curve. The reason is that third order derivatives are used in the conditions, such that all supporting points influence all pieces of the piecewise polynomial. 

Other interpolation methods lead to the same results. Namely, Lagrange interpolation polynomial \cite{Abramowitz84} and the inversion of the Vandermonde matrix \cite{Macon58} (often used in polynomial fitting or polyfit \cite{Zimichev2022}) lead to the exact same interpolation if four supporting points are used. 

In the literature, it was observed that this kind of interpolation can lead to the famous Runge's phenomenon \cite{Runge1901}, which means that the curve does not follow a smooth, monotonous behavior but that in certain cases, it shows considerable oscillations. The reason is that the constraints on the second and third order derivative can lead to overshoots. As a consequence, interpolated curves highly deviate from observable points. Hence, this method is unsuitable for practical use \cite{Strom21,Herglotz22b}. Eventually, in recent years, the primal CSI interpolation approach was replaced by PCHIP \cite{Strom21}, which is discussed in the next subsection.

\subsection*{2) Piecewise Cubic Hermite Interpolation Polynomial}
\label{secsec:PCHIP}
Piecewise Cubic Hermite Interpolation Polynomial (PCHIP) \cite{Fritsch84} was specifically designed to suppress the unwanted effect of oscillations (Runge's phenomenon). Therefore, special constraints were derived mathematically ensuring monotony when the input supporting points are monotonous, such that overshoots are suppressed. 

\begin{figure}
\psfrag{006}[tc][tc]{ }
\psfrag{007}[bc][bc]{ }
\psfrag{000}[ct][ct]{ $P_{\mathrm{ind},i-1}$}%
\psfrag{001}[ct][ct]{ $P_{\mathrm{ind},i}$}%
\psfrag{002}[ct][ct]{ $P_{\mathrm{ind},i+1}$}%
\psfrag{003}[r][r]{ $p_{\mathrm{dep},i-1}$}%
\psfrag{004}[r][r]{ $p_{\mathrm{dep},i}$}%
\psfrag{005}[r][r]{ $p_{\mathrm{dep},i+1}$}%
\psfrag{A}[c][c]{ $h_{i-1}$}%
\psfrag{B}[c][c]{ $h_{i}$}%
\psfrag{C}[c][c]{ $v_{i}$}%
\psfrag{D}[c][c]{ $\Delta _{i}$}%
\psfrag{E}[c][c]{ $\Delta _{i-1}$}%
\psfrag{Support points}[l][l]{\small{Support points}}%
\psfrag{Horizontal Distance}[l][l]{\small{Horizontal distance}}%
\psfrag{Lin. interpolation}[l][l]{\small{Slopes between points}}%
\psfrag{Tangent line}[l][l]{\small{Tangent at $P_{\mathrm{ ind},i}$}}%
\psfrag{Interpolation}[l][l]{\small{Interpolation $\hat p_\mathrm{dep}$ \eqref{eq:piecewisePolynomial}}}%
\includegraphics[width=0.5\textwidth]{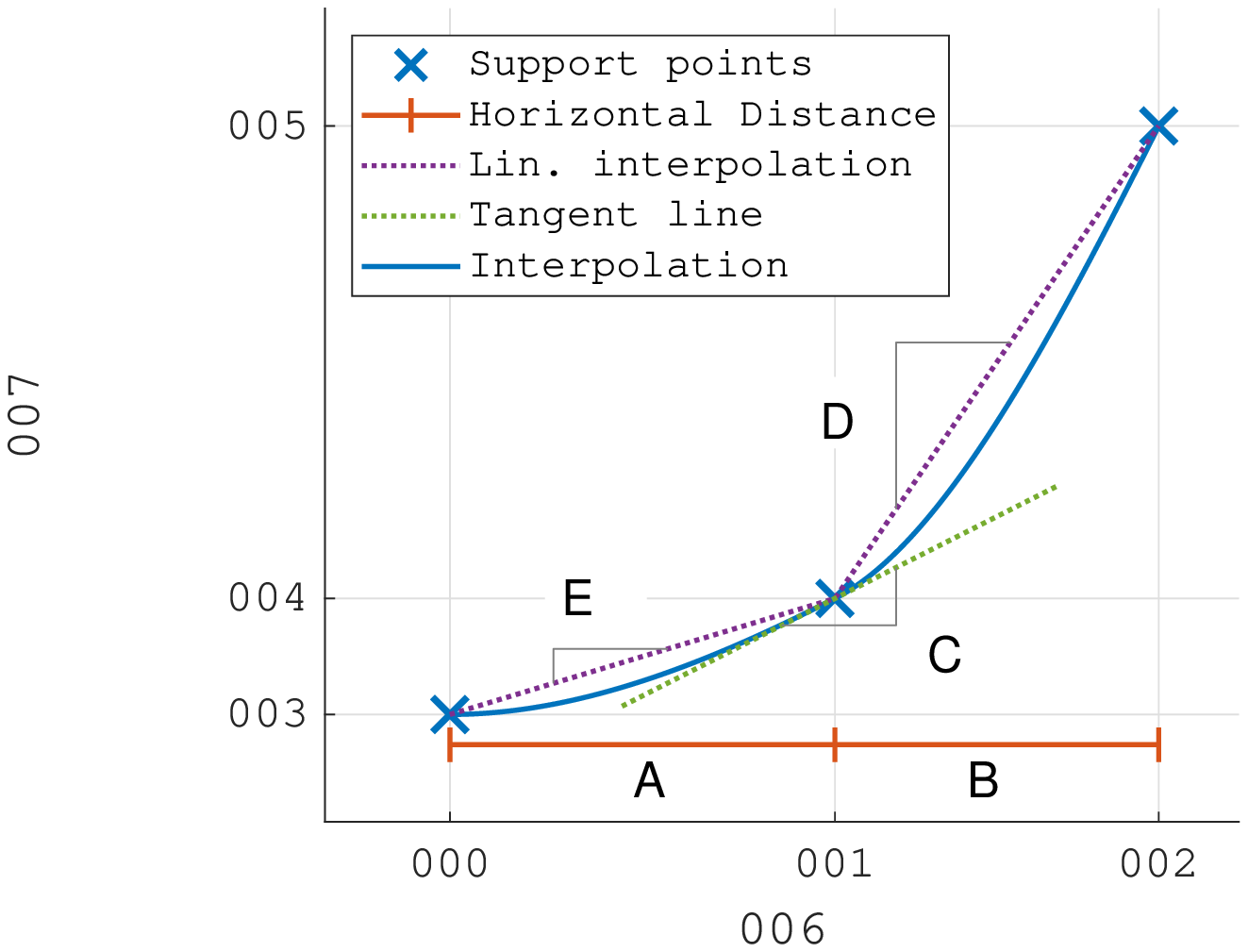}
\vspace{-0.6cm}
\caption{Calculation of the derivatives $v_i$ in the supporting points. Note that for better visualization of slopes, we swapped the horizontal and the vertical axis (i.e. the axis for the dependent and the independent variable). To calculate $v_i$, the distance between the supporting points $h_i$ and the linear slopes $\Delta_i$ between the supporting points are used. } 
\label{fig:varsForDervi}
\vspace{-.6cm}
\end{figure}

For PCHIP, the general idea is to determine suitable derivatives $v_i$ in the supporting points, which are then used to construct the remaining $I$ conditions complementing Eqs.~\eqref{eq:interpoConstBase} and \eqref{eq:slopeConstr}. If these derivatives are given, we can use the distance between two supporting points $h_i=P_{\mathrm{ind},i+1}-P_{\mathrm{ind},i}$ and the slope $\Delta_i = \frac{p_{\mathrm{dep},i+1}-p_{\mathrm{dep},i}}{h_i}$ of the linear line segment joining the two supporting points to calculate each interval of the piecewise interpolated curve by \cite{Fritsch1980} 
\begin{align}
\hat p_{\mathrm{dep},i}(\phi) = & \quad  p_{\mathrm{dep},i} + v_i(\phi - P_{\mathrm{ind},i}) \nonumber \\ 
&  +  \left(\frac{3\Delta_i -2v_i -v_{i+1}}{h_i}\right) (\phi - P_{\mathrm{ind},i})^2 \nonumber \\
& + \left(\frac{v_i+v_{i+1}-2\Delta_i}{h_i^2}\right)(\phi - P_{\mathrm{ind},i})^3.  \label{eq:piecewisePolynomial}
\end{align}
The meaning of all the variables is visualized in Fig.~\ref{fig:varsForDervi}.

The remaining question is how to determine the derivatives $v_i$ in such a way that monotony is preserved. In \cite{Fritsch84}, the PCHIP method, which was first proposed in \cite{Butland1980}, is discussed in detail and the following solution is proposed for internal supporting points ($i=2,3,...,I-1$)
\begin{equation}
v_i = \begin{cases}	
\frac{\Delta_{i-1}\Delta_i}{\alpha \Delta_i + (1-\alpha) \Delta_{i-1}}, & \Delta_{i-1}\Delta_i > 0 \\
0, & \mathrm{otherwise}, 
\end{cases}
\label{eq:PCHIPderivatives}
\end{equation}
where 
\begin{equation}
\alpha = \frac{h_{i-1}+2h_i}{3(h_{i-1}+ h_i)} \label{eq:PCHIPalpha}. 
\end{equation}
This means that if the supporting points are not monotonously increasing, an extremum is enforced at corresponding supporting points. 
For the two outer supporting points $i\in\{1,I\}$, the following equations are used
\begin{align}
v_1 =&\frac{(2h_1 + h_2)\Delta_1 - h_1 \Delta_2}{h_1 + h_2}\nonumber \\
 v_I =&\frac{(2h_I + h_{I-1})\Delta_I - h_I \Delta_{I-1}}{h_I + h_{I-1}}. \label{eq:PCHIPboundaries}
\end{align}

To interpret the meaning of these equations, we simplify by taking the assumption that the distance between the supporting points is constant, i.e., $h_{i-1} = h_i$. In this case, Eqs.~\eqref{eq:PCHIPderivatives} and \eqref{eq:PCHIPboundaries} simplify to 
\begin{align}
v_1 =&\frac{3\Delta_1 - \Delta_2}{2}\nonumber \\
v_{2,...,I-1} = & \begin{cases}	
\frac{2\Delta_{i-1}\Delta_i}{\Delta_i + \Delta_{i-1}}, & \Delta_{i-1}\Delta_i > 0 \\
0, &\mathrm{otherwise}
\end{cases} \label{eq:PCHIPderivativesSimple}\\
v_I =&\frac{3\Delta_I - \Delta_{I-1}}{2}. 
\nonumber
\end{align} 
The term for the inner supporting points corresponds to the harmonic mean of the slopes on the left $\Delta_{i-1}$ and the right $\Delta_i$ of the supporting points.

As a consequence, the term $\alpha$ can be interpreted as a correction factor if the interval sizes $h_i$ are different. Furthermore, the harmonic mean indicates that the final derivative used for interpolation will have a bias towards the lower value of the two slopes. 

\subsection*{3) Akima Interpolation}
\label{secsec:Akima}

For Akima interpolation \cite{Akima70}, we also determine the derivatives to apply Eq.~\eqref{eq:piecewisePolynomial}. The derivatives $v_i, i=1,...,I$ are calculated by 
\begin{equation}
v_i = \frac{\left| \Delta_{i+1}-\Delta_i\right|\Delta_{i-1}  + \left| \Delta_{i-1}-\Delta_{i-2}\right|\Delta_{i}}      {\left| \Delta_{i+1}-\Delta_i\right| + \left| \Delta_{i-1}-\Delta_{i-2}\right|}.  \label{eq:AkimaDerivatives}
\end{equation}
From this calculation, we can see that additional slopes $\Delta$  must be calculated, which may be beyond the limits of the supporting points. The calculation of these slopes is given by
\begin{equation}
\Delta_i = \begin{cases}
2\Delta_{i+1} - \Delta_{i+2}, & i=-1,0 \\
2\Delta_{i-1} - \Delta_{i-2}, & i= I+1, I+2. 
\end{cases}
\label{eq:AkimaAddSlopes}
\end{equation}
We can see that in contrast to PCHIP, not only the adjacent slopes, but also the slopes next to the adjacent slopes have an impact on the final interpolation. In general, a weighted linear interpolation of the adjacent slopes $\Delta_ {i-1}$ and $\Delta_i$ is performed, where the weight of each slope is higher if the slope difference on the opposite side of the current support point ($\left|\Delta_{i+1}-\Delta_i\right|$ and $\left|\Delta_{i-1}-\Delta_{i-2}\right|$, respectively) is high. Hence, Akima interpolation is biased towards homogeneous slopes and not towards smaller slopes, as was identified for PCHIP. 



\input{appendix.tex}

\input{appendix_proof_for_log.tex}

\end{document}

%% file: datasets/Matthias/RAT.tex
newly proposed tools are assessed to decide whether they should be adopted. 
Therefore, it is important to evaluate the compression gains of a proposed coding tool accurately since the improvement can be below 1\%. In the tool reporting procedure of JVET~\cite{JVET_T0013}, 
it is explained how to evaluate whether a coding tool improves the compression efficiency of the reference encoder. 
This is done by switching the usage of a specific coding tool on and off. In this work, we evaluate five distinct coding tools with respect to the default encoding configuration. 

We use VVenC v.1.5~\cite{VVenCRepo}, which is a practical encoder implementation by Fraunhofer HHI~\cite{VVenC}, 
with the slower preset as a reference. 
Based on this preset, we separately disable the following five coding tools: 
intra subpartitions (ISP), sample adaptive offset (SAO), luma mapping with chroma scaling (LMCS), bi-directional optical flow (BDOF), and adaptive loop filter (ALF). For each of these six configurations, we encode the sequences from class A1-F as suggested by the JVET CTCs for SDR content~\cite{JVET_T2010} with the random access and the lowdelay\_B test condition. Therefore, we encode the QP values from 22 to 37 with the supporting points of 22, 27, 32, and 37. As PMs, we evaluate the PSNR, bitrate, VMAF, and decoding energy demand. To derive the decoding energy demand, we use the setup that is presented in~\cite{Kraenzler21}, which uses an Intel i7-8700 CPU and utilizes its integrated power meter Running Average Power Limit (RAPL). As a decoder, we use the optimized decoder implementation VVdeC~\cite{VVdeCRepo} v1.5, which is also developed by Fraunhofer HHI~\cite{VVdeC}. In addition, the decoder energy measurements are validated with a confidence interval test.

%% file: appendix.tex
\section*{B. Considerations on Averaging}
\label{app:integr}
This appendix discusses properties and implications of the averaging procedure in the BD calculus. As mentioned in Section~III, the mean relative difference between the curves is calculated by Eq.~(6). 
 For simplification, we replace the dependent and the independent variables by $Y$ and $x$, respectively, where the latter is given in the logarithmic domain. In most cases, $Y$ represents a quality metric and $x$ the logarithm of the rate. The BD-integration then reads
\begin{equation}
\Delta X =  10^{\frac{1}{\Delta Y } \int_{Y_\mathrm{low}}^{Y_\mathrm{high}}
 x_\mathrm{A}(Y) 
 - x_\mathrm{B}(Y) \mathrm{d}Y} -1.  
 \label{eq:baseInt}
 \end{equation}
 In this equation, it is not trivial to see that a mean relative difference is calculated. Hence, we rewrite Eq. \eqref{eq:baseInt} to 
\begin{align}
\Delta X = &  10^{\frac{1}{\Delta Y } \left[ \int_{Y_\mathrm{low}}^{Y_\mathrm{high}}
 x_\mathrm{A}(Y)\mathrm{d}Y 
 - \int_{Y_\mathrm{low}}^{Y_\mathrm{high}} x_\mathrm{B}(Y) \mathrm{d}Y\right]} -1 \nonumber\\
 = & \frac{10^{\frac{1}{\Delta Y } \int_{Y_\mathrm{low}}^{Y_\mathrm{high}}
 x_\mathrm{A}(Y)\mathrm{d}Y }}{10^{\frac{1}{\Delta Y }\int_{Y_\mathrm{low}}^{Y_\mathrm{high}} x_\mathrm{B}(Y) \mathrm{d}Y}}-1, 
 \end{align}
 where we can see that the mean rate over the interval $[Y_\mathrm{low}, Y_\mathrm{high}]$ is calculated for both curves in the logarithmic domain and then put into relation by division. Subtracting $1$ ensures that the resulting value can be interpreted as a percentage. 
 
This calculus is different from calculating an average in the linear domain. Averaging in the linear domain would lead to 
 \begin{equation}
\Delta X_\mathrm{lin} =   \frac{1}{\Delta Y } \int_{Y_\mathrm{low}}^{Y_\mathrm{high}}\frac{10^{x_\mathrm{A}(Y)}}{10^{ x_\mathrm{B}(Y)}}-1\, \mathrm{d}Y. 
 \end{equation} 
In the literature, we could not find discussions or reasons for the choice of calculating the average in the logarithmic domain. However, we find the following arguments supporting that the averaging should be performed in the logarithmic domain: 
 \begin{enumerate}
 \item Averaging in the logarithmic domain puts a stronger weight to small curve differences such that the resulting mean value is biased towards the smaller difference. Assume an example in which the difference between the two curves is $\frac{X}{Y}=10$-fold in the first half of the interval $ \Delta Y$ and $\frac{X}{Y}=1000$-fold in the second half. 
 The mean in the logarithmic domain then leads to $\Delta X = 10^{\frac{\mathrm{log}10(10) + \mathrm{log}10(1000)}{2}}-1 = 10^2 - 1 = 99$, while the average mean leads to $\Delta X_\mathrm{lin} = \frac{1000+10}{2} -1 = 505$, which is much higher. Hence, averaging in the logarithmic domain leads to more conservative values. 
 \item Algorithmically, calculating the mean in the logarithmic domain leads to a simpler implementation because the integration can be calculated as a closed form solution (integration of a simple third-oder polynomial). This is not possible in the linear domain, where an integration of a polynomial in an exponential function must be handled. 
 \end{enumerate}

%% file: appendix_proof_for_log.tex
\section*{C. Proof of the Equivalent Logarithm}
 In Section~III, it was stated that using logarithms of different bases does not lead to differences in final BD values. To prove this mathematically, we show that 
\begin{equation}
10^{\int x_{10}(Y) \mathrm{d}Y} = \mathrm{e}^{\int x_{\mathrm{e}}(Y) \mathrm{d}Y},  
\end{equation}
with 
\begin{align}
x_{10}(Y) = \mathrm{log}_{10}(X(Y)),\\
x_{\mathrm{e}}(Y) = \mathrm{ln}(X(Y)). 
\end{align}
In the RD sense, $X$ corresponds to the rate, $x$ to the rate in log-domain, and $Y$ to the distortion. 

For simplicity, we first assume that the interpolated function in $\mathrm{ln}$-domain is equal to the interpolated function in $\mathrm{log}_{10}$-domain (will be proven later). Then, we can apply the logarithm $\mathrm{log}_{10}(z) = \frac{\mathrm{ln}(z)}{\mathrm{ln}(10)}$ to both sides as 
\begin{align}
\int x_{10}(Y) \mathrm{d}Y = \frac{1}{\mathrm{ln}(10)}\int x_{\mathrm{e}}(Y) \mathrm{d}Y
\end{align}
and further solve as 
\begin{align}
\int \mathrm{log}_{10}(X(Y))) \mathrm{d}Y & = \int \frac{1}{\mathrm{ln}(10)}\mathrm{ln}(X(Y)) \mathrm{d}Y \nonumber\\
& = \int \mathrm{log}_\mathrm{10}(X(Y)) \mathrm{d}Y, 
\end{align}
which proves our initial assumption. 

Concerning the equality of the interpolation, we find that in the supporting points, the values of the dependent variables in the logarithmic domains (basis $10$ and basis e) differs by a constant factor. For example, given the value $x_\mathrm{e}$ in the natural-logarithm domain, the factor is  $\alpha= \frac{1}{\mathrm{ln}(10)}$ as 
\begin{equation}
x_\mathrm{10} = \log_{10}(X) = \frac{\mathrm{ln}(X)}{\mathrm{ln}(10)} = \alpha \cdot x_\mathrm{e}. 
\end{equation}
Consequently, the interpolated functions in both logarithmic domains return the $\alpha$-scaled value in the supporting points. 

Concerning the constraints on the derivatives from Eqs. (8)-(11), 
the linear factor leads to the same result in both domains because derivatives adhere to the property of linearity. 

Finally, the constraints on the derivatives in the supporting points  $v_i$ for PCHIP and Akima interpolation (Eqs. \eqref{eq:PCHIPderivatives}, \eqref{eq:PCHIPboundaries},\eqref{eq:AkimaDerivatives}), 
also return the same result because they are calculated by linear combinations of the slopes $\Delta_i$ between the supporting points. Consequently, the interpolated curves from the natural logarithm and the logarithm to the basis $10$ are equivalent. 

\section*{D. Further Datasets}

This appendix explains additional datasets evaluated for the study in this paper. The datasets are summarized in Table~II.

\subsection{Codec Comparison {(CC)}}
\input{datasets/Matthias/CodecComparison.tex}

\subsection{Encoder Preset Evaluation {(ENC)}}
\input{datasets/Geetha_ENC/dataset_description.tex}

\subsection{AOM Common Test Conditions {(AOM)}}
\input{datasets/Lena_AV1/dataset_description.tex}


\subsection{Neural-Network based Video Compression \emph{(NNV)}}
\input{datasets/Anna_LearnedVideo/learnedVideoCompression.tex}

\begin{table*}[t]
\renewcommand{\arraystretch}{1.3}
\caption{RIEs and subset errors for LDT data set.  }
\label{tab:LDTerrors}
\vspace{-0.3cm}
\begin{center}
\begin{tabular}{l|r|r|r|r|r|r}
\hline
& \multicolumn{2}{c|}{PSNR - bitrate} & \multicolumn{2}{c|}{VMAF - Rate} & \multicolumn{2}{c}{log VMAF - Rate} \\
& $\bar e^\mathrm{RIE}$ & $E_\mathrm{max}^\mathrm{RIE}$ & $\bar e^\mathrm{RIE}$ & $E_\mathrm{max}^\mathrm{RIE}$ & $\bar e^\mathrm{RIE}$ & $E_\mathrm{max}^\mathrm{RIE}$ \\ \hline
PCHIP & \small $0.902\%$ & \small $18.703\%$  & \small $4.692\%$ & \small $79.704\%$  & \small $1.500\%$ & \small $48.786\%$ \\ 
Akima & \small $0.798\%$ & \small $18.131\%$  & \small $3.773\%$ & \small $72.534\%$  & \small $1.417\%$ & \small $49.690\%$ \\ 
\hline
$\bar e^\mathrm{sub} | e_\sigma^\mathrm{sub}$ & \small $0.074\%$ & \small $0.124\%$  & \small $0.153\%$ & \small $0.274\%$  & \small $0.243\%$ & \small $0.489\%$   
\\ 
\hline 
\end{tabular}
\end{center}
\end{table*}

\begin{table*}[t]
\renewcommand{\arraystretch}{1.3}
\caption{RIEs and subset errors for CC data set.  }
\label{tab:CCerrors}
\vspace{-0.3cm}
\begin{center}
\begin{tabular}{l|r|r|r|r|r|r}
\hline
& \multicolumn{2}{c|}{PSNR-bitrate} & \multicolumn{2}{c|}{VMAF-bitrate}& \multicolumn{2}{c}{log VMAF-bitrate}\\
& $\bar e^\mathrm{RIE}$ & $E_\mathrm{max}^\mathrm{RIE}$ & $\bar e^\mathrm{RIE}$ & $E_\mathrm{max}^\mathrm{RIE}$ & $\bar e^\mathrm{RIE}$ & $E_\mathrm{max}^\mathrm{RIE}$\\ \hline
PCHIP & \small $0.489\%$ & \small $8.762\%$  & \small $3.509\%$ & \small $68.064\%$  & \small $1.684\%$ & \small $48.243\%$ \\ 
Akima & \small $0.479\%$ & \small $8.443\%$  & \small $2.782\%$ & \small $66.407\%$  & \small $1.604\%$ & \small $46.885\%$ \\ 
\hline
$\bar e^\mathrm{sub} | e_\sigma^\mathrm{sub}$ & \small $0.462\%$ & \small $0.323\%$  & \small $1.210\%$ & \small $1.154\%$  & \small $1.164\%$ & \small $1.612\%$  \\ 
\hline
\end{tabular}
\end{center}
\end{table*}

\begin{table*}[ht]
\renewcommand{\arraystretch}{1.3}
\caption{RIEs and subset errors for the ENC data set. }
\label{tab:x265errors}
\vspace{-0.3cm}
\begin{center}
\resizebox{\textwidth}{!}{ 
\begin{tabular}{l|r|r|r|r|r|r|r|r|r|r|r|r|r|r}
\hline
& \multicolumn{2}{c|}{PSNR-bitrate} & \multicolumn{2}{c|}{SSIM-bitrate}& \multicolumn{2}{c|}{log SSIM-bitrate} & \multicolumn{2}{c|}{PSNR-Enc. Energy} & \multicolumn{2}{c|}{ SSIM-Enc. Energy}& \multicolumn{2}{c|}{log SSIM-Enc. Energy} & \multicolumn{2}{c}{PSNR-Enc. Time} \\
& $\bar e^\mathrm{RIE}$ & $E_\mathrm{max}^\mathrm{RIE}$ & $\bar e^\mathrm{RIE}$ & $E_\mathrm{max}^\mathrm{RIE}$ & $\bar e^\mathrm{RIE}$ & $E_\mathrm{max}^\mathrm{RIE}$& $\bar e^\mathrm{RIE}$ & $E_\mathrm{max}^\mathrm{RIE}$& $\bar e^\mathrm{RIE}$ & $E_\mathrm{max}^\mathrm{RIE}$ & $\bar e^\mathrm{RIE}$ & $E_\mathrm{max}^\mathrm{RIE}$ & $\bar e^\mathrm{RIE}$ & $E_\mathrm{max}^\mathrm{RIE}$  \\ \hline
PCHIP & \small $0.594\%$ & \small $8.686\%$  & \small $1.641\%$ & \small $16.558\%$  & \small $0.672\%$ & \small $8.969\%$  & \small $0.538\%$ & \small $7.266\%$  & \small $0.721\%$ & \small $9.279\%$  & \small $0.555\%$ & \small $6.811\%$  & \small $0.350\%$ & \small $6.501\%$ \\ 
Akima & \small $0.539\%$ & \small $8.253\%$  & \small $1.152\%$ & \small $13.511\%$  & \small $0.617\%$ & \small $7.582\%$  & \small $0.539\%$ & \small $6.665\%$  & \small $0.661\%$ & \small $8.194\%$  & \small $0.557\%$ & \small $6.332\%$  & \small $0.343\%$ & \small $5.978\%$ \\ 
\hline
$\bar e^\mathrm{sub} | e_\sigma^\mathrm{sub}$& \small $0.340\%$ & \small $0.450\%$  & \small $0.959\%$ & \small $0.702\%$  & \small $0.357\%$ & \small $0.448\%$  & \small $5.942\%$ & \small $13.082\%$  & \small $8.598\%$ & \small $18.252\%$  & \small $5.370\%$ & \small $11.260\%$  & \small $3.966\%$ & \small $7.846\%$ \\ 
\hline
\end{tabular}}
\end{center}
\end{table*}

\begin{table*}[ht]
\renewcommand{\arraystretch}{1.3}
\caption{RIEs and subset errors for AOM data set. }
\label{tab:vp9AV1errors}
\vspace{-0.3cm}
\begin{center}
\resizebox{\textwidth}{!}{ 
\begin{tabular}{l|r|r|r|r|r|r|r|r|r|r}
\hline
& \multicolumn{2}{c|}{PSNR-bitrate} & \multicolumn{2}{c|}{PSNR-Enc. Time} & \multicolumn{2}{c|}{log SSIM-bitrate} & \multicolumn{2}{c|}{log MS-SSIM-bitrate} & \multicolumn{2}{c}{log VMAF-bitrate}\\
& $\bar e^\mathrm{RIE}$ & $E_\mathrm{max}^\mathrm{RIE}$ & $\bar e^\mathrm{RIE}$ & $E_\mathrm{max}^\mathrm{RIE}$ & $\bar e^\mathrm{RIE}$ & $E_\mathrm{max}^\mathrm{RIE}$ & $\bar e^\mathrm{RIE}$ & $E_\mathrm{max}^\mathrm{RIE}$ & $\bar e^\mathrm{RIE}$ & $E_\mathrm{max}^\mathrm{RIE}$ \\ \hline
PCHIP & \small $0.531\%$ & \small $10.299\%$  & \small $4.640\%$ & \small $77.179\%$  & \small $0.735\%$ & \small $8.549\%$  & \small $0.751\%$ & \small $8.782\%$  & \small $1.396\%$ & \small $19.237\%$ \\ 
Akima & \small $0.519\%$ & \small $12.705\%$  & \small $4.660\%$ & \small $73.191\%$  & \small $0.847\%$ & \small $22.676\%$  & \small $0.794\%$ & \small $13.595\%$  & \small $1.533\%$ & \small $32.622\%$ \\ 
\hline
$\bar e^\mathrm{sub} | e_\sigma^\mathrm{sub}$ & \small $2.267\%$ & \small $3.848\%$  & \small $0.113\%$ & \small $0.176\%$  & \small $2.827\%$ & \small $5.318\%$  & \small $3.923\%$ & \small $8.064\%$  & \small $3.846\%$ & \small $6.815\%$ \\ 
\hline
 \end{tabular}}
\end{center}
\end{table*}

\subsection{Screen Content Coding {SCC}}
\input{datasets/Hannah_SCC/screenContentCoding.tex}

\subsection{Video Coding for Machines ({VCM})}
\input{datasets/Kristian_VCM/dataset_description.tex}

\subsection{$360^\circ$-Video (360)}
\input{datasets/Andy_360/dataset_description.tex}

\begin{table*}[ht]
\renewcommand{\arraystretch}{1.3}
\caption{RIEs and subset errors for NNV data set. }
\label{tab:NNvid_errors}
\vspace{-0.3cm}
\begin{center}
\begin{tabular}{l|r|r|r|r|r|r|r|r}
\hline
& \multicolumn{2}{c|}{PSNR-YUV-bitrate} & \multicolumn{2}{c|}{PSNR-RGB-bitrate}& \multicolumn{2}{c|}{log MSSSIM-bitrate} & \multicolumn{2}{c}{log VMAF-bitrate}  \\
& $\bar e^\mathrm{RIE}$ & $E_\mathrm{max}^\mathrm{RIE}$ & $\bar e^\mathrm{RIE}$ & $E_\mathrm{max}^\mathrm{RIE}$ & $\bar e^\mathrm{RIE}$ & $E_\mathrm{max}^\mathrm{RIE}$ & $\bar e^\mathrm{RIE}$ & $E_\mathrm{max}^\mathrm{RIE}$ \\ \hline
PCHIP & \small $0.401\%$ & \small $6.015\%$  & \small $0.371\%$ & \small $5.520\%$  & \small $0.551\%$ & \small $19.341\%$  & \small $1.048\%$ & \small $14.813\%$ \\ 
Akima & \small $0.420\%$ & \small $6.758\%$  & \small $0.390\%$ & \small $8.958\%$  & \small $0.586\%$ & \small $27.871\%$  & \small $1.135\%$ & \small $16.411\%$ \\ 
\hline
$\bar e^\mathrm{sub} | e_\sigma^\mathrm{sub}$ & \small $0.484\%$ & \small $0.959\%$  & \small $0.467\%$ & \small $0.713\%$   & \small $1.008\%$ & \small $2.140\%$  & \small $1.616\%$ & \small $2.884\%$ \\ 
\hline
\end{tabular}
\end{center}
\end{table*}

\begin{table*}[ht]
\renewcommand{\arraystretch}{1.3}
\caption{RIEs and subset errors for SCC data set.  }
\label{tab:SCCerrors}
\vspace{-0.3cm}
\begin{center}
\begin{tabular}{l|r|r|r|r|r|r|r|r|r|r}
\hline
& \multicolumn{2}{c|}{PSNR RGB-bitrate} & \multicolumn{2}{c|}{log VMAF$_\mathrm{YUV444}$-bitrate} & \multicolumn{2}{c|}{ESIM-bitrate} & \multicolumn{2}{c|}{GFM-bitrate} & \multicolumn{2}{c}{GMSD-bitrate}\\
& $\bar e^\mathrm{RIE}$ & $E_\mathrm{max}^\mathrm{RIE}$ & $\bar e^\mathrm{RIE}$ & $E_\mathrm{max}^\mathrm{RIE}$ & $\bar e^\mathrm{RIE}$ & $E_\mathrm{max}^\mathrm{RIE}$ & $\bar e^\mathrm{RIE}$ & $E_\mathrm{max}^\mathrm{RIE}$ & $\bar e^\mathrm{RIE}$ & $E_\mathrm{max}^\mathrm{RIE}$ \\ \hline
PCHIP & \small $0.251\%$ & \small $1.003\%$  & \small $0.681\%$ & \small $3.808\%$  & \small $0.312\%$ & \small $1.442\%$  & \small $0.852\%$ & \small $4.111\%$  & \small $1.033\%$ & \small $4.503\%$ \\ 
Akima & \small $0.258\%$ & \small $0.979\%$  & \small $0.513\%$ & \small $3.409\%$  & \small $0.273\%$ & \small $1.338\%$  & \small $0.489\%$ & \small $2.225\%$  & \small $0.648\%$ & \small $2.498\%$ \\ 
\hline
$\bar e^\mathrm{sub} | e_\sigma^\mathrm{sub}$ & \small $0.125\%$ & \small $0.101\%$  & \small $0.267\%$ & \small $0.199\%$  & \small $0.117\%$ & \small $0.131\%$  & \small $0.110\%$ & \small $0.131\%$  & \small $0.258\%$ & \small $0.077\%$ \\ 
\hline
 \end{tabular}
\end{center}
\end{table*}

\begin{table*}[ht]
\renewcommand{\arraystretch}{1.3}
\caption{RIEs and subset errors for VCM data set (QP $\in \{22,27,32,37\}$).  }
\label{tab:VCM_high_errors}
\vspace{-0.3cm}
\begin{center}
\resizebox{\textwidth}{!}{ 
\begin{tabular}{l|r|r|r|r|r|r|r|r|r|r|r|r|r|r|r|r|r|r}
\hline
highQP &  \multicolumn{2}{c|}{DeepLab} & \multicolumn{4}{c|}{FasterRCNN} & \multicolumn{4}{c|}{MaskRCNN} & \multicolumn{4}{c|}{Yolov5}&   \multicolumn{2}{c|}{DeepLab} &\multicolumn{2}{c}{Yolov5} \\
& \multicolumn{2}{c|}{mIoU-bitrate} & \multicolumn{2}{c|}{mAP-bitrate} & \multicolumn{2}{c|}{wAP-bitrate} & \multicolumn{2}{c|}{mAP-bitrate} & \multicolumn{2}{c|}{wAP-bitrate} & \multicolumn{2}{c|}{mAP-bitrate} & \multicolumn{2}{c|}{wAP-bitrate}& \multicolumn{2}{c|}{log mIoU-bitrate} & \multicolumn{2}{c}{log wAP-bitrate}\\
& $\bar e^\mathrm{RIE}$ & $E_\mathrm{max}^\mathrm{RIE}$ & $\bar e^\mathrm{RIE}$ & $E_\mathrm{max}^\mathrm{RIE}$ & $\bar e^\mathrm{RIE}$ & $E_\mathrm{max}^\mathrm{RIE}$ & $\bar e^\mathrm{RIE}$ & $E_\mathrm{max}^\mathrm{RIE}$& $\bar e^\mathrm{RIE}$ & $E_\mathrm{max}^\mathrm{RIE}$& $\bar e^\mathrm{RIE}$ & $E_\mathrm{max}^\mathrm{RIE}$& $\bar e^\mathrm{RIE}$ & $E_\mathrm{max}^\mathrm{RIE}$& $\bar e^\mathrm{RIE}$ & $E_\mathrm{max}^\mathrm{RIE}$& $\bar e^\mathrm{RIE}$ & $E_\mathrm{max}^\mathrm{RIE}$\\ \hline
PCHIP & \small $1.329\%$ & \small $4.888\%$  & \small $3.336\%$ & \small $12.449\%$  & \small $1.559\%$ & \small $7.140\%$  & \small $3.368\%$ & \small $20.255\%$  & \small $1.724\%$ & \small $7.734\%$  & \small $8.398\%$ & \small $27.354\%$  & \small $2.865\%$ & \small $12.239\%$  & \small $1.186\%$ & \small $4.208\%$  & \small $2.853\%$ & \small $12.238\%$ \\ 
Akima & \small $1.236\%$ & \small $4.082\%$  & \small $3.166\%$ & \small $11.185\%$  & \small $1.489\%$ & \small $6.438\%$  & \small $2.832\%$ & \small $19.120\%$  & \small $1.588\%$ & \small $6.769\%$  & \small $8.096\%$ & \small $27.324\%$  & \small $2.746\%$ & \small $12.244\%$  & \small $1.211\%$ & \small $4.222\%$  & \small $2.750\%$ & \small $12.243\%$ \\ 
\hline
$\bar e^\mathrm{sub} | e_\sigma^\mathrm{sub}$ &\small $1.763\%$ & \small $0.000\%$  & \small $2.317\%$ & \small $0.000\%$  & \small $2.239\%$ & \small $0.000\%$  & \small $2.970\%$ & \small $0.000\%$  & \small $1.380\%$ & \small $0.000\%$  & \small $13.451\%$ & \small $0.000\%$  & \small $2.002\%$ & \small $0.000\%$  & \small $1.498\%$ & \small $0.000\%$  & \small $2.046\%$ & \small $0.000\%$ \\ 
\hline 
 \end{tabular}}
\end{center}
\end{table*}

\begin{table*}[t]
\renewcommand{\arraystretch}{1.3}
\caption{RIEs and subset errors for VCM data set (QP $\in \{12,17,22,27\}$). }
\label{tab:VCM_low_errors}
\vspace{-0.3cm}
\begin{center}
\resizebox{\textwidth}{!}{ 
\begin{tabular}{l|r|r|r|r|r|r|r|r|r|r|r|r|r|r|r|r|r|r}
\hline
lowQP &  \multicolumn{2}{c|}{DeepLab} & \multicolumn{4}{c|}{FasterRCNN} & \multicolumn{4}{c|}{MaskRCNN} & \multicolumn{4}{c|}{Yolov5}&   \multicolumn{2}{c|}{DeepLab} &\multicolumn{2}{c}{Yolov5} \\
& \multicolumn{2}{c|}{mIoU-bitrate} & \multicolumn{2}{c|}{mAP-bitrate} & \multicolumn{2}{c|}{wAP-bitrate} & \multicolumn{2}{c|}{mAP-bitrate} & \multicolumn{2}{c|}{wAP-bitrate} & \multicolumn{2}{c|}{mAP-bitrate} & \multicolumn{2}{c|}{wAP-bitrate}& \multicolumn{2}{c|}{log mIoU-bitrate} & \multicolumn{2}{c}{log wAP-bitrate}\\
& $\bar e^\mathrm{RIE}$ & $E_\mathrm{max}^\mathrm{RIE}$ & $\bar e^\mathrm{RIE}$ & $E_\mathrm{max}^\mathrm{RIE}$ & $\bar e^\mathrm{RIE}$ & $E_\mathrm{max}^\mathrm{RIE}$ & $\bar e^\mathrm{RIE}$ & $E_\mathrm{max}^\mathrm{RIE}$& $\bar e^\mathrm{RIE}$ & $E_\mathrm{max}^\mathrm{RIE}$& $\bar e^\mathrm{RIE}$ & $E_\mathrm{max}^\mathrm{RIE}$& $\bar e^\mathrm{RIE}$ & $E_\mathrm{max}^\mathrm{RIE}$& $\bar e^\mathrm{RIE}$ & $E_\mathrm{max}^\mathrm{RIE}$& $\bar e^\mathrm{RIE}$ & $E_\mathrm{max}^\mathrm{RIE}$\\ \hline
PCHIP & \small $4.388\%$ & \small $24.132\%$  & \small $13.681\%$ & \small $82.572\%$  & \small $6.675\%$ & \small $26.349\%$  & \small $9.008\%$ & \small $41.859\%$  & \small $5.247\%$ & \small $27.168\%$  & \small $14.783\%$ & \small $100.000\%$  & \small $14.288\%$ & \small $86.952\%$  & \small $4.218\%$ & \small $24.408\%$  & \small $14.313\%$ & \small $87.117\%$ \\ 
Akima & \small $3.119\%$ & \small $12.018\%$  & \small $5.889\%$ & \small $24.648\%$  & \small $6.403\%$ & \small $24.678\%$  & \small $8.473\%$ & \small $24.248\%$  & \small $4.694\%$ & \small $27.048\%$  & \small $9.944\%$ & \small $45.680\%$  & \small $6.741\%$ & \small $48.027\%$  & \small $3.020\%$ & \small $11.887\%$  & \small $6.735\%$ & \small $48.019\%$ \\ 
\hline
$\bar e^\mathrm{sub} | e_\sigma^\mathrm{sub}$ & \small $1.157\%$ & \small $0.000\%$  & \small $9.590\%$ & \small $0.000\%$  & \small $4.461\%$ & \small $0.000\%$  & \small $0.822\%$ & \small $0.000\%$  & \small $3.875\%$ & \small $0.000\%$  & \small $3.101\%$ & \small $0.000\%$  & \small $11.862\%$ & \small $0.000\%$  & \small $1.507\%$ & \small $0.000\%$  & \small $11.441\%$ & \small $0.000\%$ \\ 
\hline
 \end{tabular}}
\end{center}
\end{table*}

\subsection{Point Cloud Compression {(PCG and PCA)}}
\input{datasets/PointCloudCoding_Dat/pointcloudcoding}

\section*{E. Interpolation and Subset Errors}
\label{app:interpErrorTables}

This appendix lists all mean relative interpolation errors  (RIEs) and subset errors for all tested data sets and performance metrics in Tables~\ref{tab:LDTerrors} to \ref{tab:360errors}.

\begin{table*}[t]
\renewcommand{\arraystretch}{1.3}
\caption{RIEs and subset errors for 360 data set.  }
\label{tab:360errors}
\vspace{-0.3cm}
\begin{center}
\resizebox{\textwidth}{!}{ 
\begin{tabular}{l|r|r|r|r|r|r|r|r|r|r|r|r|r|r}
\hline
& \multicolumn{2}{c|}{Y-E2ESPSNR-NN-filesize} & \multicolumn{2}{c|}{U-E2ESPSNRN-N-filesize} & \multicolumn{2}{c|}{V-E2ESPSNR-NN-filesize} & \multicolumn{2}{c|}{Y-E2EWSPSNR-filesize} & \multicolumn{2}{c|}{Y-PSNR-DYN-VP0-filesize} & \multicolumn{2}{c|}{Y-CFSPSNR-NN-filesize} & \multicolumn{2}{c}{Y-CFCPPPSNR-filesize}\\
& $\bar e^\mathrm{RIE}$ & $E_\mathrm{max}^\mathrm{RIE}$ & $\bar e^\mathrm{RIE}$ & $E_\mathrm{max}^\mathrm{RIE}$ & $\bar e^\mathrm{RIE}$ & $E_\mathrm{max}^\mathrm{RIE}$ & $\bar e^\mathrm{RIE}$ & $E_\mathrm{max}^\mathrm{RIE}$& $\bar e^\mathrm{RIE}$ & $E_\mathrm{max}^\mathrm{RIE}$& $\bar e^\mathrm{RIE}$ & $E_\mathrm{max}^\mathrm{RIE}$ & $\bar e^\mathrm{RIE}$ & $E_\mathrm{max}^\mathrm{RIE}$ \\ \hline
PCHIP & \small $0.792\%$ & \small $3.796\%$  & \small $2.103\%$ & \small $12.539\%$  & \small $2.054\%$ & \small $11.921\%$  & \small $0.771\%$ & \small $3.899\%$  & \small $0.945\%$ & \small $5.070\%$  & \small $0.791\%$ & \small $3.776\%$  & \small $0.788\%$ & \small $3.922\%$ \\ 
Akima & \small $0.608\%$ & \small $3.045\%$  & \small $2.255\%$ & \small $14.401\%$  & \small $2.131\%$ & \small $12.159\%$  & \small $0.585\%$ & \small $3.060\%$  & \small $0.775\%$ & \small $4.892\%$  & \small $0.606\%$ & \small $3.040\%$  & \small $0.610\%$ & \small $3.097\%$ \\ 
\hline
$\bar e^\mathrm{sub} | e_\sigma^\mathrm{sub}$ & \small $0.311\%$ & \small $0.338\%$  & \small $0.692\%$ & \small $0.902\%$  & \small $0.817\%$ & \small $1.041\%$  & \small $0.302\%$ & \small $0.280\%$  & \small $0.243\%$ & \small $0.299\%$  & \small $0.310\%$ & \small $0.337\%$  & \small $0.285\%$ & \small $0.285\%$ \\ 
\hline
 \end{tabular}}
\end{center}
\end{table*}

\begin{table*}[t]
\renewcommand{\arraystretch}{1.3}
\caption{RIEs and subset errors for PCA data set. For Cb-PSNR-bitrate and Cr-PSNR-bitrate, only a subset of sequences was evaluated because the performance curves did not overlap. }
\label{tab:PCC_attr_errors}
\vspace{-0.3cm}
\begin{center}
\begin{tabular}{l|r|r|r|r|r|r}
\hline
& \multicolumn{2}{c|}{Y-PSNR-bitrate} & \multicolumn{2}{c|}{Cb-PSNR-bitrate} & \multicolumn{2}{c}{Cr-PSNR-bitrate} \\
& $\bar e^\mathrm{RIE}$ & $E_\mathrm{max}^\mathrm{RIE}$ & $\bar e^\mathrm{RIE}$ & $E_\mathrm{max}^\mathrm{RIE}$ & $\bar e^\mathrm{RIE}$ & $E_\mathrm{max}^\mathrm{RIE}$ \\ \hline
PCHIP & \small $0.452\%$ & \small $5.143\%$  & \small $0.802\%$ & \small $14.778\%$  & \small $1.327\%$ & \small $31.247\%$ \\ 
Akima & \small $0.499\%$ & \small $5.066\%$  & \small $0.834\%$ & \small $14.787\%$  & \small $1.314\%$ & \small $32.660\%$ \\ 
\hline
$\bar e^\mathrm{sub} | e_\sigma^\mathrm{sub}$ & \small $0.646\%$ & \small $0.620\%$  &  \small $21.468\%$ & \small $53.785\%$ & \small $1.315\%$ & \small $1.831\%$ \\ 
\hline
 \end{tabular}
\end{center}
\end{table*}

\begin{table*}[t]
\renewcommand{\arraystretch}{1.3}
\caption{RIEs and subset errors for PCG data set. }
\label{tab:PCC_geom_errors}
\vspace{-0.3cm}
\begin{center}
\begin{tabular}{l|r|r|r|r}
\hline
& \multicolumn{2}{c|}{d1-PSNR-bitratePerInputPoint} & \multicolumn{2}{c}{d2-PSNR-bitratePerInputPoint}\\
& $\bar e^\mathrm{RIE}$ & $E_\mathrm{max}^\mathrm{RIE}$ & $\bar e^\mathrm{RIE}$ & $E_\mathrm{max}^\mathrm{RIE}$ \\
\hline
PCHIP & \small $2.733\%$ & \small $40.269\%$  & \small $2.794\%$ & \small $53.075\%$ \\
Akima & \small $2.432\%$ & \small $30.973\%$  & \small $2.753\%$ & \small $49.797\%$  \\
\hline
$\bar e^\mathrm{sub} | e_\sigma^\mathrm{sub}$ & \small $3.231\%$ & \small $4.322\%$  & \small $5.360\%$ & \small $3.400\%$ \\ 
\hline
 \end{tabular}
\end{center}
\end{table*}

%% file: datasets/Matthias/CodecComparison.tex

In the Codec Comparison {(CC)} data set, we evaluate the compression efficiency of the two video reference encoder implementations HM 16.26~\cite{HM} (HEVC) and VTM 17.2~\cite{VTM} (VVC). For both, we use the sequences from class B, C, D, and F of the JVET CTC~\cite{JVET_T2010}. Furthermore, we use the randomaccess test conditions to encode the first 128 frames of each sequence with all QPs from 22 to 37. As supporting points, we use the QP values 22, 27, 32, and 37. As performance metrics, we use PSNR, bitrate, and VMAF.

%% file: datasets/Geetha_ENC/dataset_description.tex
Many encoders provide the possibility to set the desired encoding complexity with a so-called `preset', where rate-distortion performance is traded with encoding time. With this data set, we evaluate the performance of three presets  \textit{veryfast}, \textit{medium}, and  \textit{slower} of two different encoders in terms of compression efficiency and encoding energy. 
We generate these data with the x265 \cite{x265} and the x264 encoder implementation \cite{x264} that perform multi-core processing on an Intel Xeon CPU. We consider 22 sequences from the JVET CTCs \cite{JVET_N1010} 
and encode the first 64 frames of each sequence at the mentioned presets and crf values $\in \{18, 19, ..., 33\}$, with four subset supporting points $18,23,28,33$. 
We obtain the encoder processing energy using the energy measurement procedure in \cite{Herglotz18}, which was already  explained for the TOO data set. 
Furthermore, the performance metrics encoding time, bitrate, PSNR, SSIM, and VMAF values are also recorded.

%% file: datasets/Lena_AV1/dataset_description.tex
The data sets introduced before all used four supporting points for $\BD$ calculations. This data set is based on CTCs that use six supporting points \cite{AOM_CTC21}. We investigate the AOMedia Video (AV1) video codec, which achieves higher compression performance than its predecessor VP9~\cite{Mukherjee13}
at the cost of additional computational complexity. 
We use the Scalable Video Technology (SVT) implementation, that achieves a better trade-off regarding quality 
and speed~\cite{Kossentini20}, to evaluate the compression performance of both AV1 and VP9 encoders in terms of their 
compression performance and encoding time. 
We obtain the metrics encoding eight sequences from Class A of the AOM CTCs~\cite{AOM_CTC21} 
with presets 4, 6, and 9 and randomaccess configuration. The crf setting ranges 
from $17$ to $42$, where six supporting points are suggested for the BD calculations: 
$\left\{ 17, 22, 27, 32, 37, 42\right\}$.
The encoding time is measured as the sum of user and system time on single core execution, the rate is measured in 
kbps, and to evaluate the quality, PSNR, SSIM, MS-SSIM, and VMAF are used.

%% file: datasets/Anna_LearnedVideo/learnedVideoCompression.tex
To assess neural-network based video compression (NNV), we choose DMC \cite{Li2022}, which is the first learned video compression method to outperform VTM for YUV444 input. It is based on conditional coding and uses a hybrid spatial-temporal entropy model. DMC supports flexible rate adjustment in a single model via content-adaptive quantization. For testing, we use the publicly available I- and P-frame model\footnote{https://github.com/microsoft/DCVC/tree/main/ACMMM2022} optimized for MSE with an intra-period of 32. We use the four quantization step (QS) values determined during training as supporting points. DMC employs different QS values for the latent space of the I-frame and P-frame model as well as the motion vectors. The QS values for the P-frame latents are $1.541$, $1.083$, $0.729$, and $0.500$, for example. As intermediate points, we use two interpolated values in between each supporting point ($1.312$, $1.117$, $0.951$, $0.810$, $0.690$, and $0.587$ in the previous example). Thus, we obtain ten rate points from a single model.  

For $\BD$ calculations, we use VTM 17.2 and HM 16.25 lowdelay-P with an intra-period of 32, since DMC does not support B-frames. Concerning the video data set, we evaluate the performance of DMC on the widely used data sets UVG \cite{Mercat2020} and MCL-JCV \cite{Wang2016}. The raw 8-bit videos have a resolution of $1920 \times 1080$ each and we test on the first $96$ frames. While HM and VTM can handle the sequences' YUV420 color format, the test videos are converted to RGB before being processed by DMC. Concerning quality metrics, we measure the bitrate in bits per pixel (bpp). For quality evaluation, we use PSNR and MS-SSIM \cite{Wang2003} in the RGB domain, which is common in neural-network video compression. Additionally, we provide VMAF scores \cite{VMAF}. 


%% file: datasets/Hannah_SCC/screenContentCoding.tex
We evaluate the performance of different screen content coding tools for the compression of screen content images using VTM-17.2~\cite{VTM}. Furthermore, we include results from HM-16.21 SCM-8.8~\cite{HM}. 
200 images in RGB 4:4:4 format from the SCI1-K test set~\cite{Yang2021} are evaluated in all-intra configuration. 
The dataset contains screen content images with resolutions $1280\times720$, $1920\times1080$, and $2560\times140$.
Each image is coded in accordance with the CTC for non-4:2:0 colour formats for computer-generated content~\cite{Chao2020} as well as the adapted configurations with the following three coding tool modifications: palette mode (PLT) off, intra block copy (IBC) off, as well as both IBC and PLT switched off. 
We measure the bitrate in bits per pixel (bpp). For quality evaluation, we calculate the commonly used PSNR and SSIM in the RGB domain. Additionally, we provide evaluation results with GMSD~\cite{Xue2014} as well as dedicated screen content metrics, namely ESIM~\cite{Ni2017} and GFM~\cite{Ni2018}, to evaluate the perceptual image quality. QPs are chosen in the range $22$ to $37$ with support points at $22$, $27$, $32$, $37$. 

%% file: datasets/Kristian_VCM/dataset_description.tex
For the use case of having a neural network as information sink in a VCM scenario, we follow the VCM evaluation framework described in~\cite{Fischer20b}, which is also inline with the CTCs for the MPEG VCM standardization~\cite{Liu2020}.
As data set, we select the Cityscapes validation data set~\cite{Cordts16} containing 500 uncompressed RGB images that capture urban scenarios in German cities with $2048\times 1024$ pixels and which are annotated for several classes such as pedestrians or cars.
The VCM investigations compare the coding performance of VTM-10.0 over its predecessor HEVC with HM-16.18 as software.
We consider two different QP ranges: first, the JVET-suggested  QP range $\mathrm{QP} \in \{22, 23, ..., 37\}$, which is denoted as \textit{high QP} range.
Second, we also consider a \textit{low QP} range $\mathrm{QP} \in \{12, 13, ..., 27\}$, resulting in a higher task accuracy of the applied analysis network, which is more relevant for practical scenarios. Corresponding support points are $\mathrm{QP} \in \{22, 27, 32, 37\}$ and $\mathrm{QP} \in \{12, 17, 22, 27\}$, respectively. 

Throughout this work, we study three different analysis tasks performed on the compressed images.
The selected models are all trained on the Cityscapes training data set.
As analysis networks, we employ self-trained Faster R-CNN~\cite{Ren15b} with ResNet50 FPN backbone and YOLO-v5~\cite{Jocher2020} models for the task of object detection.
For the task of instance segmentation, we take the Mask R-CNN model with ResNet50 FPN backbone from the Detectron2 library~\cite{Wu2019} off-the-shelf.
From the same library, we use the DeepLabV3+~\cite{Chen2018} model with R103-DC5 backbone for the task of semantic segmentation.
To evaluate their performances on the compressed input data, the tasks of object detection and instance segmentation are evaluated with the mean average precision~(mAP).
To compensate class imbalances, we also consider the weighted average precision~(wAP) as introduced in~\cite{Fischer20b}.
We evaluate the DeepLabV3+ performance with the mean intersection over union~(mIoU).
To assess all three metrics, we follow the Cityscapes evaluation code from~\cite{Cordts16}.

%% file: datasets/Andy_360/dataset_description.tex
We evaluate the performance of different projection formats for the compression of $360^\circ$ videos using VTM-17.2~\cite{VTM}.
Ten sequences from the classes S1 and S2 in the JVET test set for $360^\circ$ video~\cite{JVET_E1030} are tested in randomaccess configuration.
The original sequences are downscaled to a resolution of $4096\times2048$ for class S1 and a resolution of $3072\times1536$ for class S2.
All sequences are in equirectangular projection (ERP) format with 4:2:0 chroma subsampling.
In accordance with the CTCs for $360^\circ$ video~\cite{JVET_E1030}, each sequence is coded in ERP and hybrid equiangular cubemap (HEC) projection format at a resolution of $2216\times1108$ for the ERP and $1920\times1280$ for the HEC format.
All projection format conversions and quality metric evaluations are performed using the 360Lib software 360Lib-12.0~\cite{360Lib}. QPs are chosen in the range $\mathrm{QP} \in \{22, 23, ..., 37\}$ with support points at $\mathrm{QP} \in \{22, 27, 32, 37\}$. 

%% file: datasets/PointCloudCoding_Dat/pointcloudcoding.tex
For point cloud (PC) compression methods, we use a set of ten point clouds from MPEG CTC \cite{8i}. These testing point clouds are diverse in terms of content characteristics and geometry resolutions. Specifically, we select redandblack\_vox10\_1550.ply, loot\_vox10\_1200.ply,  longdress\_vox10\_1300.ply, and soldier\_vox10\_0690.ply from the 8i data set \cite{d20178i}, which compose of smooth surface and dense point clouds at 10-bit depth. The other 12-bit depth point clouds are from MPEG CAT1 \cite{8i}, which is a data set for cultural heritage and 3D photography applications.  PCs from this set are sparse and contain more rough surfaces.

Concerning the metrics, for attribute compression, we measure the average rate per input point for all color channels (bits per point, bpp). The quality distortion metrics used in our experiments are the peak signal-to-noise ratio in dB of each individual YCbCr component which is computed using the MPEG PCC \texttt{pc\_error} tools \cite{dmetric}. Similarly, we measure the average bitrate of the geometry component in bits per input point (bpp) and use the \texttt{pc\_error} tools to compute the D1-PSNR (point-to-point) and the D2-PSNR (point-to-plane) metrics for quantitative evaluation. 

For point cloud attribute compression (PCA), we test two well-known compression algorithms: the  G-PCC v14 \cite{8571288} and RAHT v1.1\cite{de2016compression}. RAHT (Region Adaptive Hierarchical Transform) is a geometry-dependent transform method that resembles an adaptive variation of the Haar wavelet to transform the point cloud attributes. By improving entropy coding and adding coefficients prediction, RAHT became one of the core modules in the MPEG Geometry-based PCC (G-PCC). We follow the CTC \cite{8i} to generate the test results with QPs from 22 to 51 and support points at $22$, $28$, $34$, $40$, $46$, and $51$. 

Concerning the geometry compression software (PCG), octree decomposition is the straightforward way to represent point cloud geometry and has been adopted in G-PCC (Octree-based G-PCC). G-PCC also provides a predictive coding scheme that predicts each vertex of the tree from its ancestors.  Following MPEG PCC CTC, we set up two test configurations for lossy geometry coding with and without the predictive scheme. We set the rate points  \texttt{positionQuantizationScale} (pQS) from 0 to 1 (15 steps) and leave other parameters in default. Support points are $0.03125$, $0.0625$, $0.125$, $0.25$, $0.5$, and $0.75$.